\documentclass[twocolumn,dvipsnames]{aastex62}
\pdfoutput=1 
\usepackage{amsmath,amstext}
\usepackage{apjfonts}
\usepackage[figure,figure*]{hypcap}
\usepackage{ulem}
\usepackage{xspace}

\usepackage{xfrac}
\usepackage{graphicx}
\usepackage{natbib}
\usepackage{amssymb}
\usepackage{amsmath}
\usepackage{array,booktabs}
\usepackage{mathptmx}
\usepackage{booktabs}
\usepackage{hyperref}
\usepackage{float}
\usepackage{todonotes}

\DeclareSymbolFont{cmletters}{OML}{cmm}{m}{it}
\DeclareMathSymbol{v}{\mathalpha}{cmletters}{"76}
\newcommand{\rph}{\,{R_{\rm ph}}}
\newcommand{\rtr}{\,{R_{\rm tr}}}
\newcommand{\tobs}{\,{t_{\rm obs}}}
\newcommand{\thobs}{\,{\theta_{\rm obs}}}
\newcommand{\mej}{\,{M_{\rm ej}}}
\newcommand{\msun}{\,{M_{\odot}}}
\newcommand{\Ye}{\,{Y_{\rm e}}}

\shorttitle{Early NUV/Optical Emission in BH--NS Mergers}
\shortauthors{Gottlieb et al.}

\begin{document}
\title{Hours-long Near-UV/Optical Emission from Mildly Relativistic Outflows in Black Hole--Neutron Star Mergers}

    \author[0000-0003-3115-2456]{Ore Gottlieb}
	\email{oregottlieb@gmail.com}
	\affiliation{Center for Interdisciplinary Exploration \& Research in Astrophysics (CIERA), Physics \& Astronomy, Northwestern University, Evanston, IL 60202, USA}

    \author[0009-0005-2478-7631]{Danat Issa}
	\affiliation{Center for Interdisciplinary Exploration \& Research in Astrophysics (CIERA), Physics \& Astronomy, Northwestern University, Evanston, IL 60202, USA}

    \author[0000-0003-2982-0005]{Jonatan Jacquemin-Ide}
	\affiliation{Center for Interdisciplinary Exploration \& Research in Astrophysics (CIERA), Physics \& Astronomy, Northwestern University, Evanston, IL 60202, USA}

    \author[0000-0003-4475-9345]{Matthew Liska}
    \affiliation{Institute for Theory and Computation, Harvard University, 60 Garden Street, Cambridge, MA 02138, USA; John Harvard Distinguished Science and ITC}

    \author[0000-0002-9182-2047]{Alexander Tchekhovskoy}
	\affiliation{Center for Interdisciplinary Exploration \& Research in Astrophysics (CIERA), Physics \& Astronomy, Northwestern University, Evanston, IL 60202, USA}

     \author[0000-0003-4617-4738]{Francois Foucart}
	\affiliation{Department of Physics and Astronomy, University of New Hampshire, 9 Library Way, Durham, NH 03824, USA}

    \author[0000-0002-5981-1022]{Daniel Kasen}
    \affiliation{Astronomy Department and Theoretical Astrophysics Center, University of California, Berkeley, Berkeley, CA 94720, USA}
    \affiliation{Physics Department, University of California, Berkeley, Berkeley, CA 94720, USA}
    \affiliation{Nuclear Science Division, Lawrence Berkeley National Laboratory, Berkeley, CA 94720, USA}

    \author[0000-0002-3635-5677]{Rosalba Perna}
    \affiliation{Department of Physics and Astronomy, Stony Brook University, Stony Brook, NY 11794-3800, USA}
    \affiliation{Center for Computational Astrophysics, Flatiron Institute, New York, NY 10010, USA}

    \author[0000-0001-9185-5044]{Eliot Quataert}
    \affiliation{Department of Astrophysical Sciences, Princeton University, Princeton, NJ 08544, USA}

    \author[0000-0002-3635-5677]{Brian D. Metzger}
    \affiliation{Department of Physics and Columbia Astrophysics Laboratory, Columbia University, Pupin Hall, New York, NY 10027, USA}
    \affiliation{Center for Computational Astrophysics, Flatiron Institute, New York, NY 10010, USA}
 
\begin{abstract}

The ongoing LIGO--Virgo--KAGRA observing run O4 provides an opportunity to discover new multi-messenger events, including binary neutron star (BNS) mergers such as GW170817, and the highly anticipated first detection of a multi-messenger black hole--neutron star (BH--NS) merger. While BNS mergers were predicted to exhibit early optical emission from mildly relativistic outflows, it has remained uncertain whether the BH--NS merger ejecta provides the conditions for similar signals to emerge. We present the first modeling of early near-ultraviolet/optical emission from mildly relativistic outflows in BH--NS mergers. Adopting optimal binary properties: a mass ratio of $q=2$ and a rapidly rotating BH, we utilize numerical relativity and general relativistic magnetohydrodynamic (GRMHD) simulations to follow the binary's evolution from pre-merger to homologous expansion. We use an M1 neutrino transport GRMHD simulation to self-consistently estimate the opacity distribution in the outflows and find a bright near-ultraviolet/optical signal that emerges due to jet-powered cocoon cooling emission, outshining the kilonova emission at early time. The signal peaks at an absolute magnitude of $\sim -15$ a few hours after the merger, longer than previous estimates, which did not consider the first principles-based jet launching. By late 2024, the Rubin Observatory will have the capability to track the entire signal evolution or detect its peak up to distances of $\gtrsim1$~Gpc. In 2026, ULTRASAT will conduct all-sky surveys within minutes, detecting some of these events within $ \sim 200 $~Mpc. The BH--NS mergers with higher mass ratios or lower BH spins would produce shorter and fainter signals.
  
\end{abstract}
	
\section{Introduction}\label{sec:introduction}

The binary neutron star (BNS) merger GW170817 demonstrated that, in addition to gravitational waves (GWs), BNSs also produce emission throughout the entire electromagnetic spectrum, making them promising mutli-messenger sources \citep[see][for reviews]{Nakar2019,Margutti2021}. While black hole (BH) mergers are unlikely to produce any electromagnetic counterparts to GWs \citep[see, however,][]{Perna2016}, the disruption of an NS during a coalescence with a BH may give rise to kilonova emission powered by radioactive decay of heavy $r$-process elements \citep[e.g.,][]{Rosswog2005,Surman2008,Metzger2010b,Tanaka2014,Fernandez2015,Fernandez2017,Foucart2015,Kawaguchi2016,Darbha2021,Wanajo2022,Ekanger2023,Gompertz2023}, making them another potential multi-messenger events that can be detected in LIGO--Virgo--KAGRA (LVK) runs.
Understanding the electromagnetic emission in BH--NS mergers is of particular interest, since it might be the primary messenger to enable the distinction between BNS and BH--NS mergers, at least for moderate mass ratios of $ q \lesssim 3 $, high BH spin, and a relatively stiff equation of state \citep[EoS; see e.g.,][]{Tanaka2014,Yang2018,Fragione2021}.

Similar to BNS mergers, BH--NS mergers may also harbor relativistic jets that give rise to additional electromagnetic counterparts, such as a short gamma-ray burst (sGRB) and afterglow emission \citep{Paczynski1991,Mochkovitch1993,Janka1999,Etienne2012,Kiuchi2015,Paschalidis2015,Ruiz2018,Hayashi2022a,Hayashi2022b}. Unlike the kilonova signal, which can be observed from various angles, the detection of a GRB relies on the alignment of the jet with our line of sight. LVK observing run O3 detected at least one BH--NS merger, GW200115 \citep[with GW200105 being an additional controversial source;][]{Abbott2021}, for which no electromagnetic counterparts were detected \citep{Dichiara2021,Zhu2021}. Thus, if these mergers powered relativistic jets, the jet prompt emission was beamed away from Earth. Similarly, most future GW detections of BH--NS mergers are expected to fall outside the narrow beaming angle of the jet, making it unlikely to observe a coincident GW--sGRB signal. Nevertheless, the interaction between the jet and the merger ejecta could give rise to a hot, energetic cocoon that produces wide-angle emission. While the presence of ejecta is expected in certain BH--NS merger configurations of mass ratio and pre-merger BH spin \citep{Shibata2006,Shibata2007,Etienne2008,Rantsiou2008,Shibata2008,Shibata2011,Duez2010,Foucart2011,Foucart2012b,Foucart2014,Foucart2017,Foucart2018,Foucart2019,Kyutoku2011,Kyutoku2013,Kyutoku2015,Kyutoku2018,Foucart2012a,Kawaguchi2015,Brege2018,Hayashi2021,Most2021}, the amount of polar ejecta might not be sufficient to generate an energetic cocoon. Therefore, the role of cocoon emission in BH--NS mergers remains unclear.

The electromagnetic observations of GW170817 provided valuable insights into the cocoon emission, revealing two distinct processes: shock breakout emission in $\gamma$/X-rays \citep{Kasliwal2017,Gottlieb2018b} and the multi-band synchrotron emission \citep[e.g.,][]{Lazzati2018,Mooley2018a,Mooley2018b}. Although the electromagnetic data within the first $ \sim 10 $ hours after GW170817 were not available, it is expected that during that period, the cocoon also emitted radiation in the near-UV (NUV)/optical bands through multiple emission mechanisms characterized by two distinct timescales:
(i) For $t \lesssim 1$ hour, the emission is attributed to enhanced $ \beta $-decay of free neutrons \citep{Gottlieb2020c} and cooling envelope emission \citep[][]{Metzger2015,Kasliwal2017,Nakar2017,Gottlieb2018a,Piro2018}.
(ii) For $t \gtrsim 1$ hour, the dominant processes include boosted radioactive decay of heavy $r$-process elements \citep[kilonova;][]{Metzger2015,Kasliwal2017,Nakar2017,Gottlieb2018a,Piro2018,Hamidani2022}, and subsequent synchrotron emission from the $ \beta $-decay electrons \citep{Gottlieb2020c}.
Notably, none of these cocoon signals have been investigated in the context of BH--NS mergers.

In this {\it Letter}, we focus on the early NUV/optical emission, which plays a crucial role in constraining various aspects such as the fate of the jet, the distribution and composition of the ejecta, and facilitating follow-up observations \citep[however, the multiple predictions cause difficulties in discriminating the potential signals; see][]{Arcavi2018,Gottlieb2020c}. We utilize numerical relativity simulations of the pre-merger phase, which we subsequently remap onto a general relativistic magnetohydrodynamic (GRMHD) setup. By evolving the system until the homologous phase, we achieve a self-similar evolution of the outflow. Based on these numerical results, we perform the first calculations of the early NUV/optical emission that are based on the self-consistent evolution of the outflow from the pre-merger phase. These first estimates of cocoon emission in BH--NS mergers serve as valuable insights for potential detection in the LVK observing runs O4 and/or O5.

\section{Calculation method}\label{sec:setup}

We build on a numerical relativity simulation in SpEC \citep{spec} that evolves the pre-merger to 8 ms post-merger. The pre-merger setup is an aligned system of BH--NS, where the BH mass and dimensionless spin are $ M_{{\rm BH},i} = 2.7\,\msun $ and $ a_i = 0.6 $, respectively, and the NS mass is $ M_{\rm NS} = 1.35\,\msun $. The post-merger BH mass and spin are $ M_{{\rm BH},f} = 3.8\,\msun $ and $ a_f = 0.86 $.
At 8 ms after the merger, we remap the numerical relativity output to 3D GRMHD simulations using the code \textsc{h-amr} \citep{Liska2022}. We consider five post-merger magnetic field configurations, where the magnetic field depends on the mass density distribution at the time of remapping, with a cutoff at $ 5\times 10^{-4} $ of the maximum density. The maximum asymptotic Lorentz factor of the jets is set by the initial magnetization of the jets, which in turn is set by the floor values of the simulations, to be $ \sigma_0 = 150 $.
Table~\ref{tab:models} summarizes the different magnetic configurations considered in the post-merger evolution. The full details of the simulation are given in a companion paper, \citet{Gottlieb2023}.

We evolve the system for several seconds post-merger such that a significant fraction of the outflow has reached homologous expansion. Then, we post process the final snapshot of each simulation to semi-analytically calculate the predicted NUV and optical emission in the first hours after the merger. We consider adiabatic cooling emission and radioactive decay of heavy $r$-process elements from sub- and mildly relativistic outflows. Thus, we ignore contributions from elements with a Lorentz factor $ \gamma > 4 $, which have yet to reach their asymptotic velocity, and consider only elements far enough from the jet axis, $ \theta > 12^\circ $, where the cocoon maintains a self-similar structure as it evolves. Therefore, our calculation is in fact a lower limit on the cocoon emission.

	\begin{table}
		\setlength{\tabcolsep}{6.2pt}
		\centering
		\renewcommand{\arraystretch}{1.2}
		\begin{tabular}{| c | c c c | c c | }
			
                \hline
			Model & $ A $ & $ \beta_p $ & $ t_f\,[{\rm s}] $ & $ \mej\,[10^{-2}\,\msun] $ & $ t_b\,[{\rm s}] $
			\\	\hline
			$ H_0 $ & $ A = 0 $ & - & $ 8 $ & 3 & - \\ 
			$ P_w $ & $ A_\phi \propto \rho^2r^3 $ & 1000 & $ 5 $ & 3 & 0.3\\ 
			$ P_c $ & $ A_\phi \propto \rho $ & 1000 & $ 5 $ & 3 & 0.1 \\   
			$ P_s $ & $ A_\phi \propto \rho $ & 100 & $ 1.8 $ & 5 & 0.05\\   
			$ T_s $ & $ A_\theta \propto \rho $ & 1 & $ 4 $ & 4 & 4 \\ 
                \hline
		\end{tabular}
		
		\caption{
			A summary of the models' parameters. The model names stand for hydrodynamic ($H$), poloidal ($P$), or toroidal ($T$) initial magnetic fields, with the subscripts indicate the strength of the field: zero ($0$), weak ($w$), canonical ($c$), or strong ($s$). $ A $ is the vector potential, $ \beta_p $ is the characteristic gas to magnetic pressure ratio, $ t_f $ is the final time of the simulation with respect to the merger, $ M_{\rm ej} $ is the amount of unbound ejecta at the homologous phase, and $ t_b $ is the breakout time of the relativistic outflow from the disk winds.
    		}
    		\label{tab:models}
	\end{table}

\begin{figure}
    \centering
    \includegraphics[scale=0.27]{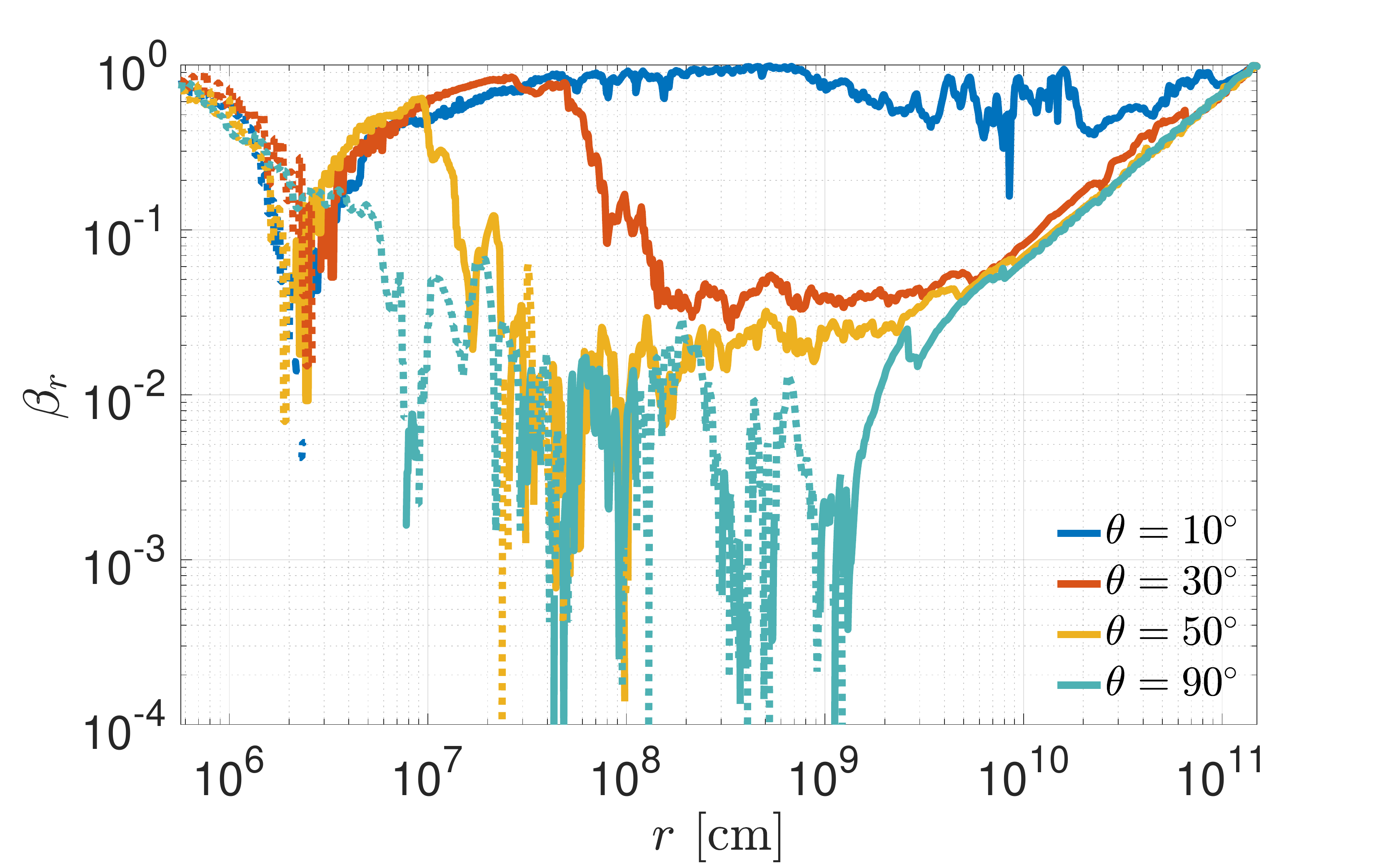}
    \includegraphics[scale=0.23]{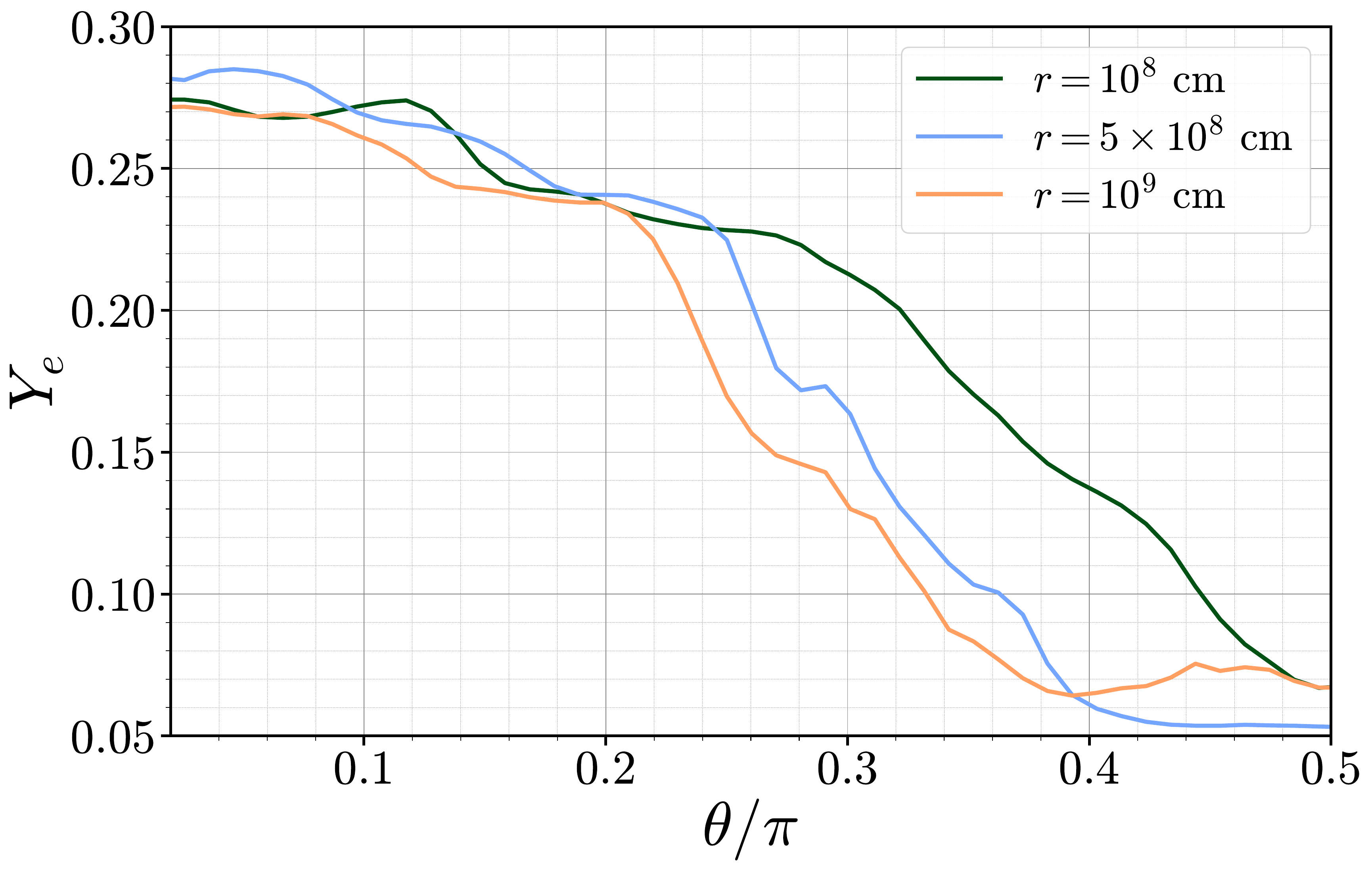}
     \caption{
     Top: dimensionless velocity profile at various polar angles, measured at $ \phi = 0 $. Solid (dotted) lines represent outflows (inflows). Homologous material ($ v \propto r $) is seen at $ r \gtrsim 10^{9.5}\,{\rm cm} $ at most angles $ \theta \gtrsim 20^\circ $. Around the jet (blue), the radial velocity profile is stochastic.
     Bottom: $\phi$-averaged $\Ye$ as a function of the polar angle at different radii, calculated from a BH--NS post-merger simulation with the M1 scheme, measured $ 0.15\,{\rm s} $ after the merger.
     }
     \label{fig:homologous}
    \end{figure}

The full description of the semi-analytic calculation is given in Appendix \ref{sec:app}. Here we summarize the main steps. At each line of sight, we consider only the gas that is homologous at this time; namely the radial velocity component dominates and scales as $ v \sim v_r \propto r $ (most of the main ejecta does not). The top panel of Figure~\ref{fig:homologous} demonstrates that the homologous expansion begins at $ r \gtrsim 10^{9.5}\,{\rm cm} $ at most angles far from the jet opening angle. At $ \theta = 10^\circ $ (blue), the flow is non-homologous at all radii due to the abrupt and turbulent jet structure. Using the radial profile, we extrapolate the gas evolution adiabatically (before it becomes radiative) to later times.

To self-consistently evolve the ejecta composition, which is probed by the electron fraction, $ \Ye $, we run an additional post-merger simulation that is identical to model $ P_c $, but employs the Helmholtz EoS and neutrino physics using M1 neutrino transport. The Helmholtz EoS \citep{Timmes_2000} is implemented in a tabulated form in $(\log \rho,\log T, \Ye)$, and includes the contributions of the ideal gas of ions, radiation and degenerate pressure of the electron--positron plasma. Neutrino transport is implemented similar to \citet{Foucart2015,Foucart2016}, with the neutrino weak force interactions included via table interpolation using NuLib \citep{OConnor_Nulib}. This simulation evolves for $ 0.15\,{\rm s} $ post-merger, while the jet is launched within the first $ 0.01\,{\rm s} $ after the merger in this model. The grid resolution is $ 128 \times 96 \times 96 $ cells. Using this simulation, we find the electron fraction distribution in the jet and disk winds to estimate $ \kappa(\theta) $ through the dependency on $ \Ye $ \citep[see, e.g.,][]{Tanaka2020}.

\begin{figure*}
    \centering
    	\includegraphics[scale=0.11]{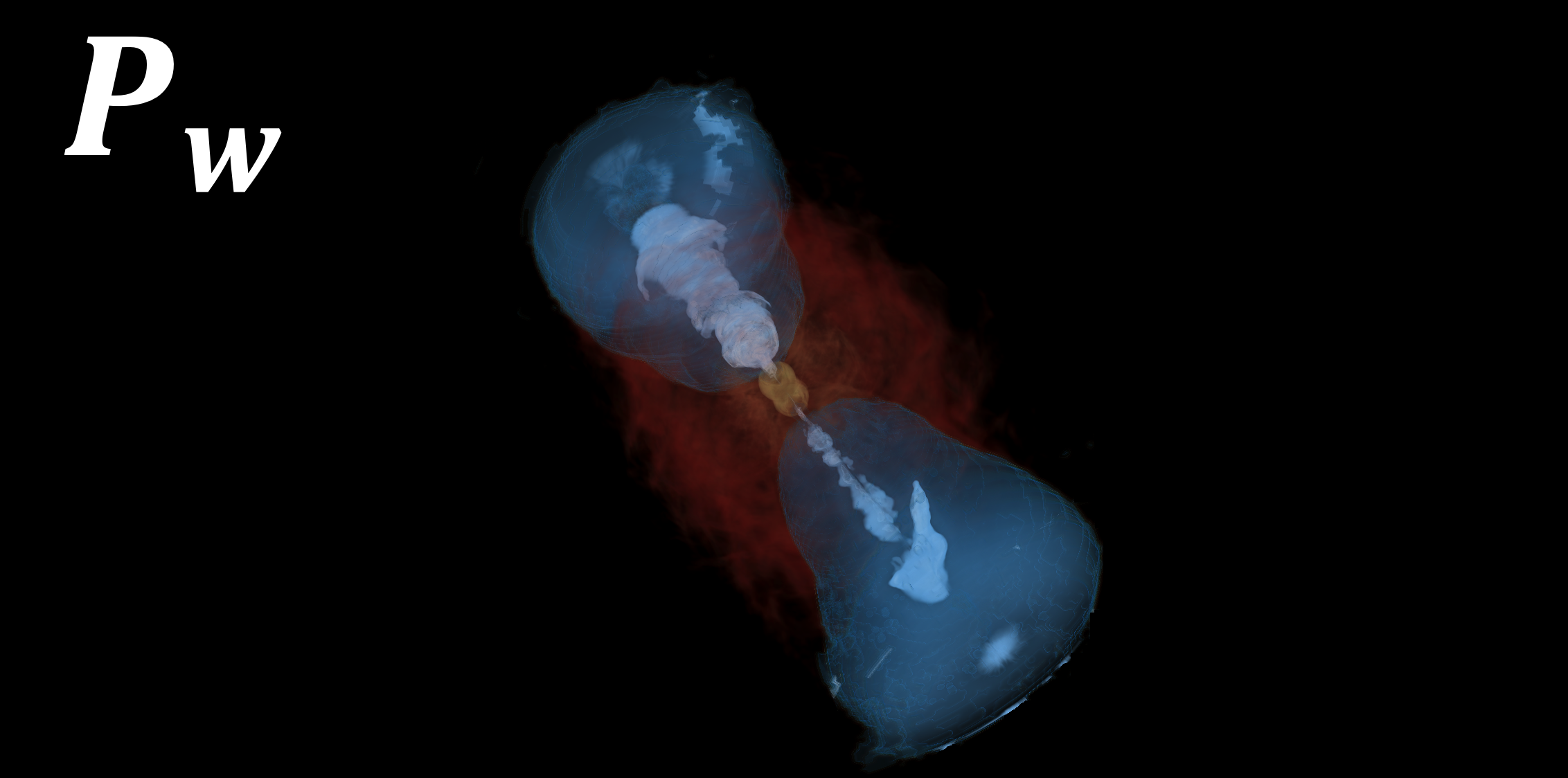}
    	\includegraphics[scale=0.11]{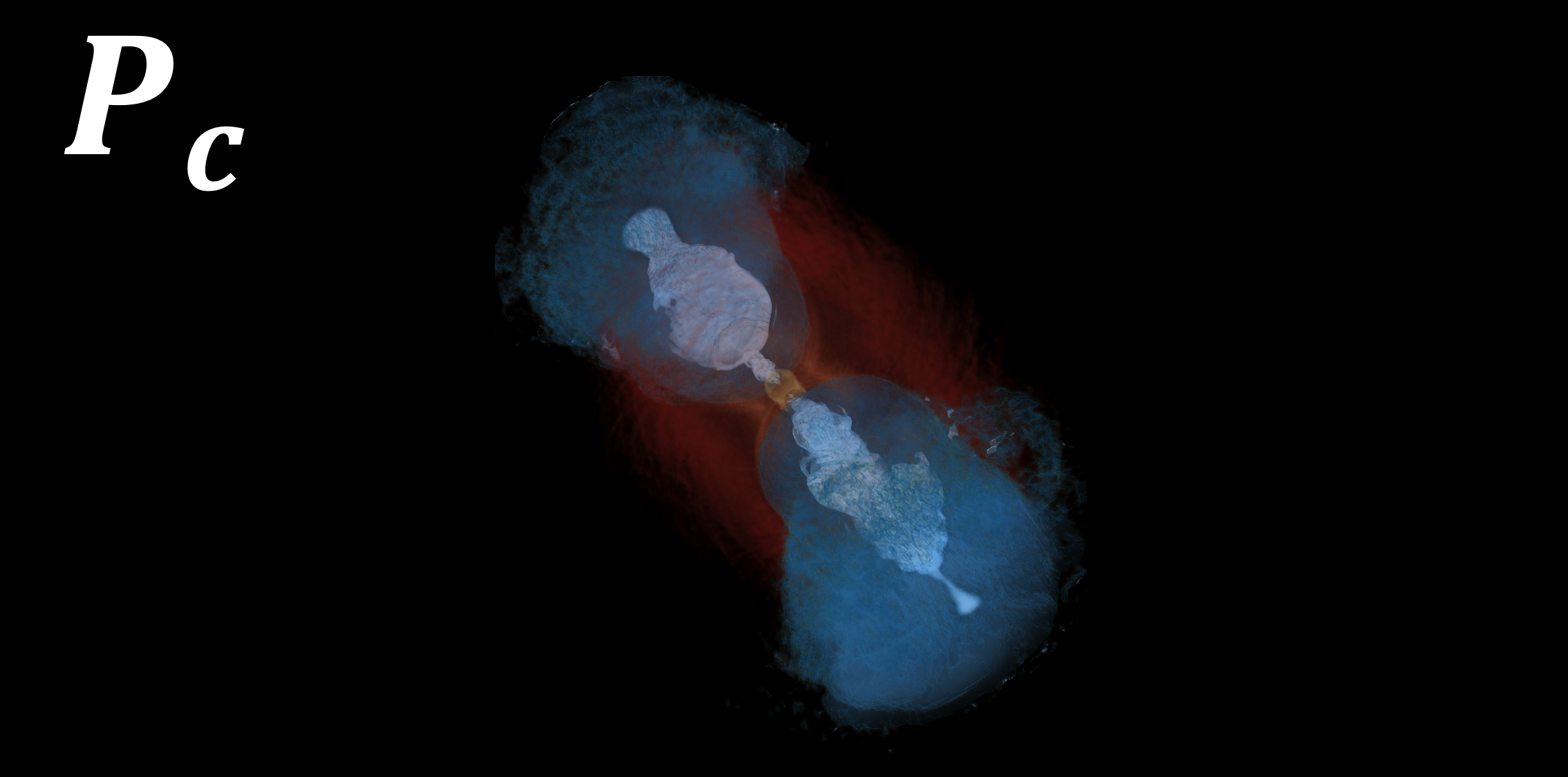}
    	\includegraphics[scale=0.11]{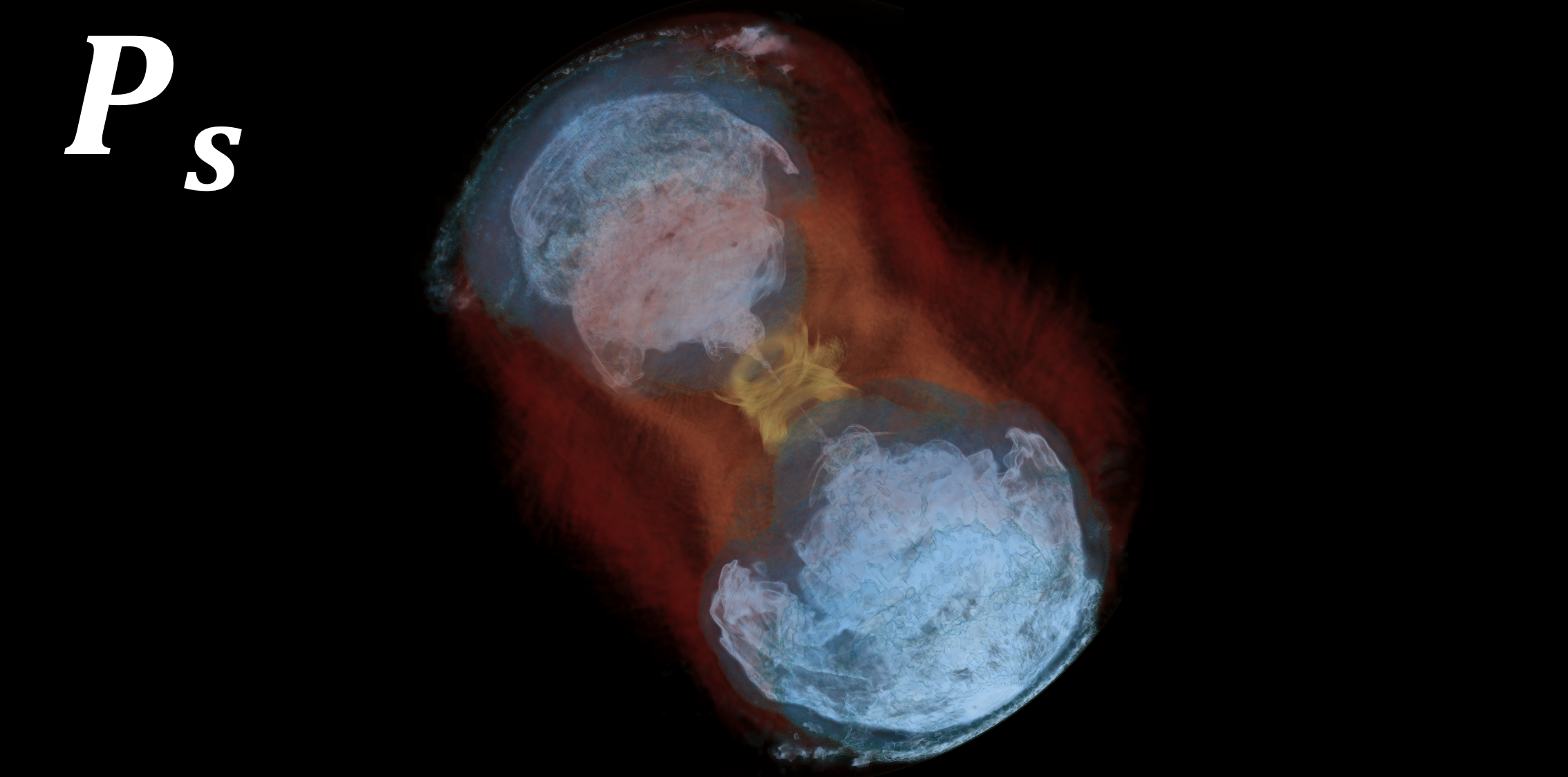}
    	\includegraphics[scale=0.11]{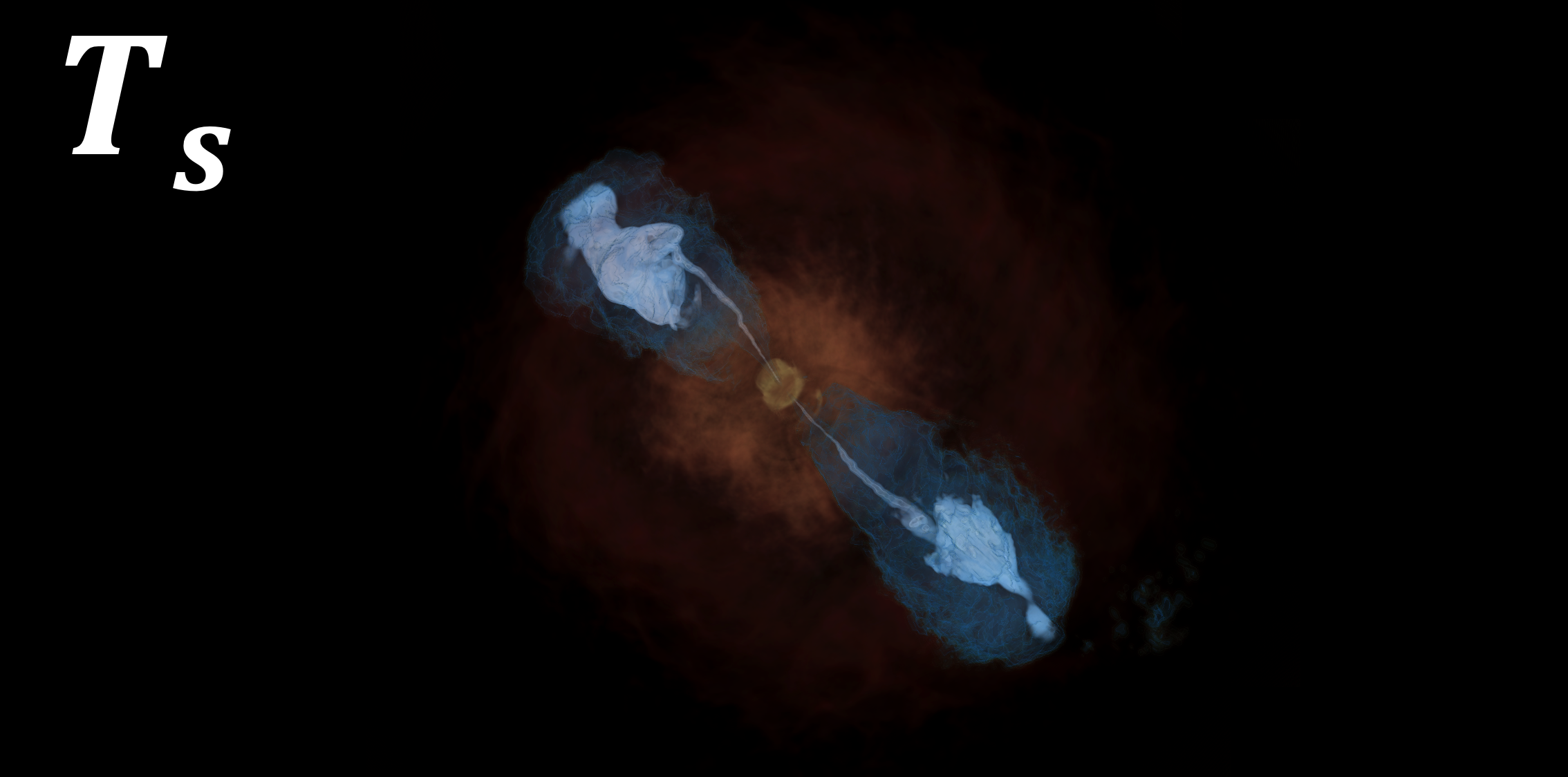}
     \caption{3D renderings of the final snapshot of each simulation. The outflows in the poloidal field configurations reach $ r \approx ct_f $, and in model $ T_s $ the outflow front is at $ r = 8.8\times 10^{10} $ cm. The blue parts display the asymptotic proper velocity, showing the sub and mildly relativistic cocoon (dark) enveloping the relativistic jets (light). The redder and yellow components portray the mass density, showing the merger ejecta and the accretion disk components, respectively. See full movies of models $ T_s $ and $ P_c $ at \url{http://www.oregottlieb.com/bhns.html}.
     }
     \label{fig:maps}
    \end{figure*}

The bottom panel of Fig.~\ref{fig:homologous} depicts the $ \phi$-averaged electron fraction as a function of the polar angle $ 0.15\,{\rm s} $ after the merger. It shows a moderate electron fraction $ \Ye \approx 0.25 $ at the jet--cocoon angles, $ \theta \lesssim 45^\circ $, which gradually drops to $ \Ye \lesssim 0.1 $ at $ \theta \gtrsim 70^\circ $. We find only a weak dependence of the electron fraction on the distance from the BH and the azimuthal angle. Thus, we use $ \Ye(\theta) $ to estimate the angle-dependent gray opacity $ \kappa(\Ye) $, as fitted by \citet{Wu2022} and consistent with \citet{Tanaka2020}:
\begin{equation}\label{eq:kappa}
    \kappa(\theta) = 1+\frac{9}{1+\left[4\Ye(\theta)\right]^{12}}\,{\rm \frac{cm^2}{g}}\,.
\end{equation}
We emphasize that the remapping of $ \Ye $ to $ \kappa $ assumes a narrow range of temperatures of several thousand Kelvin. During the first hour, the cocoon temperature is $ T \gg 10^4~{\rm K} $, for which the $r$-process opacities in the literature are only partly understood. Recent radiative transfer simulations that estimated the opacity in the range $ 10^4~{\rm K} \lesssim T \lesssim 10^5~{\rm K} $ indicate that the opacity values of high $ \Ye $ at this temperature range might be somewhat lower than in Eq.~\eqref{eq:kappa} \citep[see figure 7 in][]{Banerjee2020}, but this highly depends on the atomic composition \citep{Banerjee2023}. Therefore, we also examine the case of constant $ \kappa = 1\,{\rm cm^2\,g^{-1}} $ for comparison. In the other extreme regime of $ T \lesssim 3000~{\rm K} $, lanthanide recombination takes place and the $r$-process opacity falls dramatically \citep{Kasen2013}, but these temperatures are only of concern at later times than those studied here. Using $ \kappa(\theta) $, we calculate the radial optical depth assuming the photons diffuse radially\footnote{Deviations from sphericity were found to be important during the early BH--NS post-merger, which exhibits a quasi-planar structure, such that photons leak non-radially \citep{Kawaguchi2016,Darbha2020,Darbha2021}. However, when relativistic outflows are present, a more isotropic expanding gas structure is obtained (Fig.~\ref{fig:maps}), so that the radial diffusion approximation is reasonable.}  (Eq.~\eqref{eq:tau}). We find the trapping radius and photosphere along each line of sight and for each time.

At each time and line of sight, we calculate the local luminosity at the trapping radius and the temperature at the photosphere using the Stefan-Boltzmann law, assuming blackbody locally. We consider two emission contributions: (i) $r$-process emission $ L_r $ (Eq.~\eqref{eq:Lr}) due to radioactive heating from $ \beta $ and $ \gamma $ decay in the gas outside the trapping radius in a comoving rate $ \dot{Q}(t) \sim t^{-1.3} $ \citep[e.g.,][]{Metzger2010a} and (ii) cooling emission $ L_c $ (Eq.~\eqref{eq:Lc}) from the thermal energy of the gas and $ \beta $ and $ \gamma $ radioactive heating inside the trapping radius. Finally, we boost the bolometric and spectral luminosities to the observer frame and integrate the emission over all angles and equal light travel times.
 
\section{Outflow evolution and emission}\label{sec:results}

When the initial magnetic configuration in the disk is of a strong poloidal field, the jet power of $ \sim 10^{52}\,{\rm erg\,s^{-1}} $ is considerably higher than that of typical sGRBs. In all other models, the jet power is $ 10^{50}-10^{51}\,{\rm erg\,s^{-1}} $, consistent with typical sGRBs. However, in these cases, the jet operates for several seconds, exceeding the characteristic sub-second duration of sGRBs but consistent with long-duration sGRBs \citep[e.g.,][]{Rastinejad2022}. When the initial field is toroidal, the jet can only be launched once the dynamo process generates a global poloidal field after $ \sim 1\,{\rm s} $. Regardless of the specific magnetic configuration, turbulence in the disk releases quasi-isotropic massive winds (Table~\ref{tab:models}) during the first $ \sim 1\,{\rm s} $ after the merger \citep{Gottlieb2023}.

The interaction between the winds and jets generates a mildly relativistic hot cocoon, which is composed of shocked jet and shocked wind material, at the expense of jet power. This implies that forming a cocoon does not necessitate dynamical ejecta. Thus, cocoons accompany all jets, given that jets require an accretion disk that launches winds.
Figure~\ref{fig:maps} depicts 3D renderings of the four models with magnetic fields that generate relativistic outflows, taken at $ t = t_f $. Along the equatorial region, low $ \Ye $ winds (red) from the disk (yellow) are shocked by the relativistic jet (light blue) along the polar axis, resulting in the formation of a massive, hot cocoon (dark blue) surrounding the jet. The cocoon, consisting of $r$-process elements, emits both radioactive heating emission and cooling emission due to its adiabatic expansion.

Using a characteristic cocoon mass of $ M \approx 10^{-3}\,\msun$ and proper velocity of $\beta \approx 0.7 $, the observed emission peaks on a timescale given by Equation~\eqref{eq:tpeak}:
\begin{equation}\label{eq:time}
\begin{split}
\tobs \approx &\left[\frac{\kappa M \gamma}{4 v c}\left(1-\frac{v}{c}\right)^3\right]^{0.5} \approx \\ & 3\left(\frac{\kappa}{{\rm 5\,cm^2\,g^{-1}}}\frac{M}{10^{-3}M_\odot}\frac{\left(\frac{1-\beta}{0.3}\right)^3}{\frac{\beta}{0.7}}\right)^{0.5}\,{\rm hours}\,.
\end{split}
\end{equation}
The emission profile is expected to vary with the viewing angle due to differences in opacity (characterized by $\Ye$), mass, and velocity. Lighter, faster, and higher $ \Ye $ ejecta along the polar axis enables earlier diffusion of photons, which reduces adiabatic losses and leads to an earlier and brighter peak.

    \begin{figure*}
    \centering
    	\includegraphics[scale=0.26]{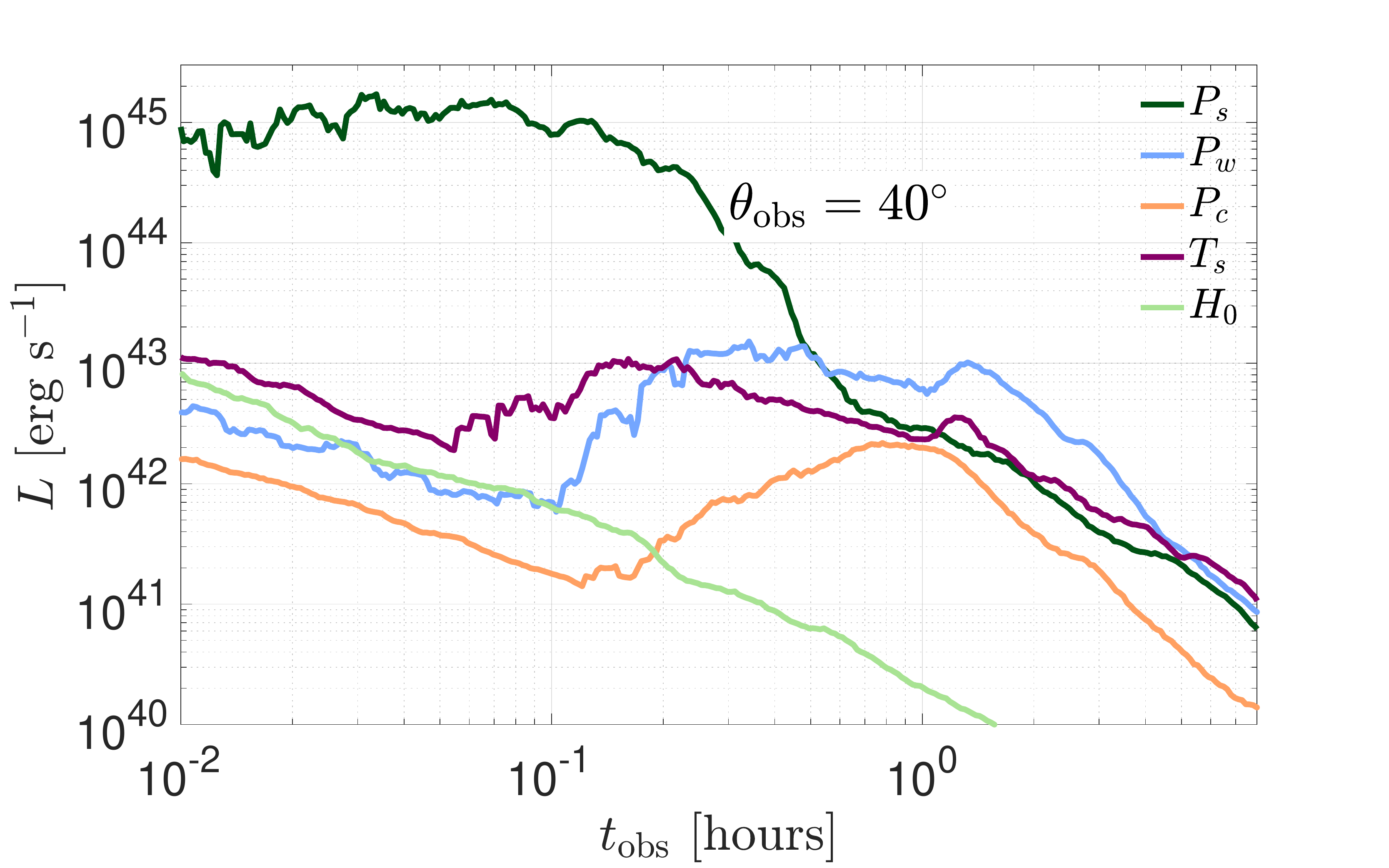}
    	\includegraphics[scale=0.26]{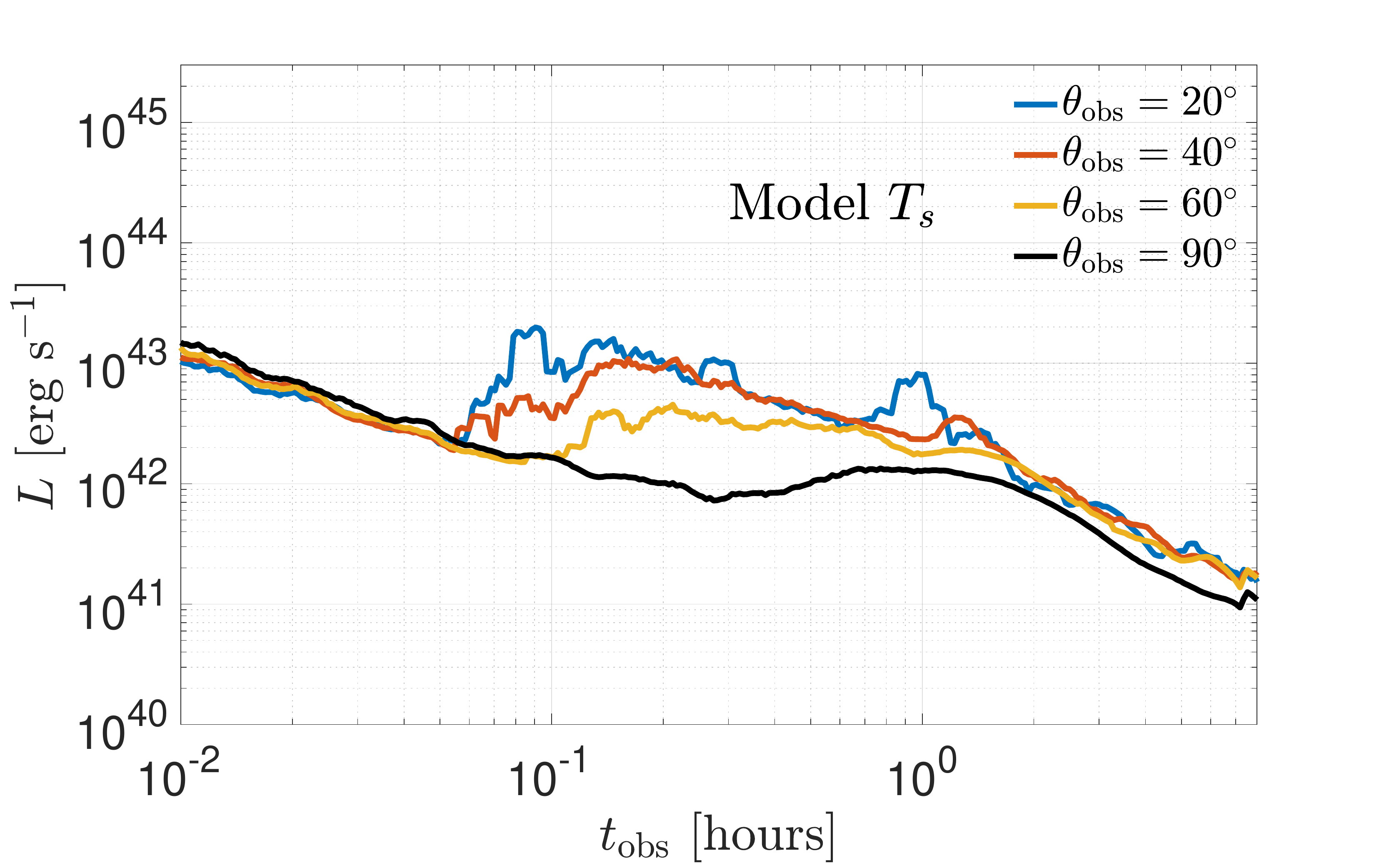}
    	\includegraphics[scale=0.26]{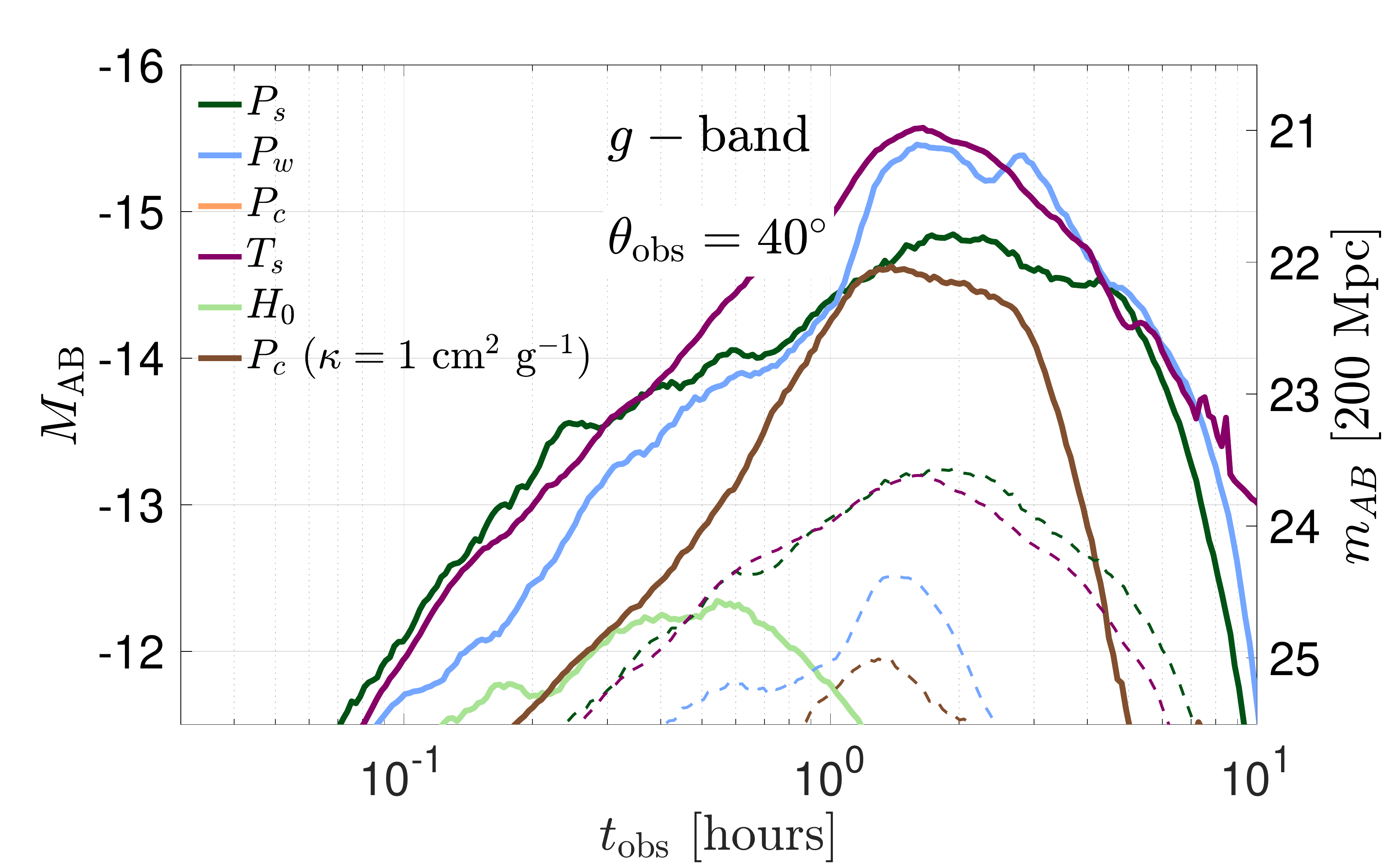}
    	\includegraphics[scale=0.26]{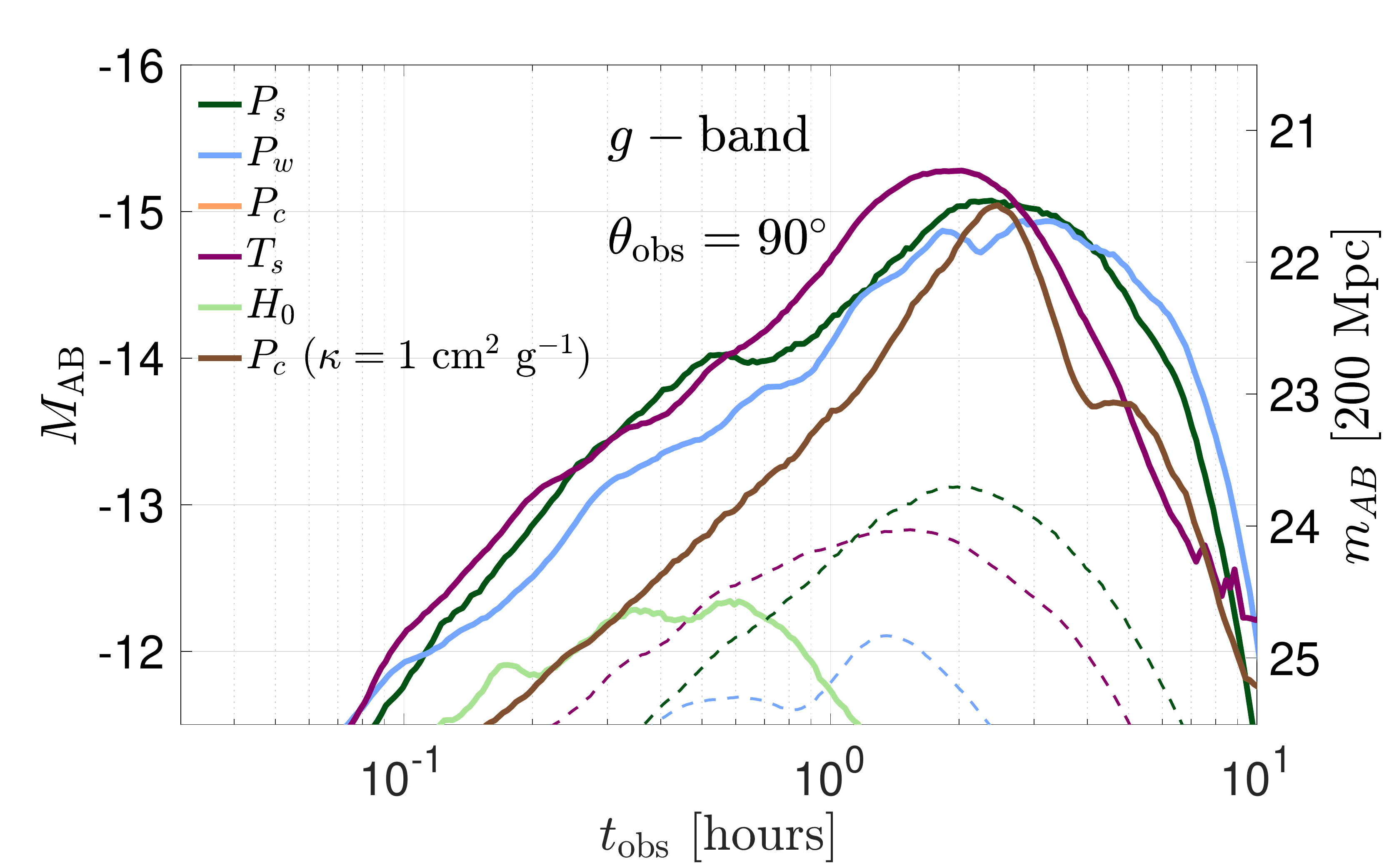}
    	\includegraphics[scale=0.26]{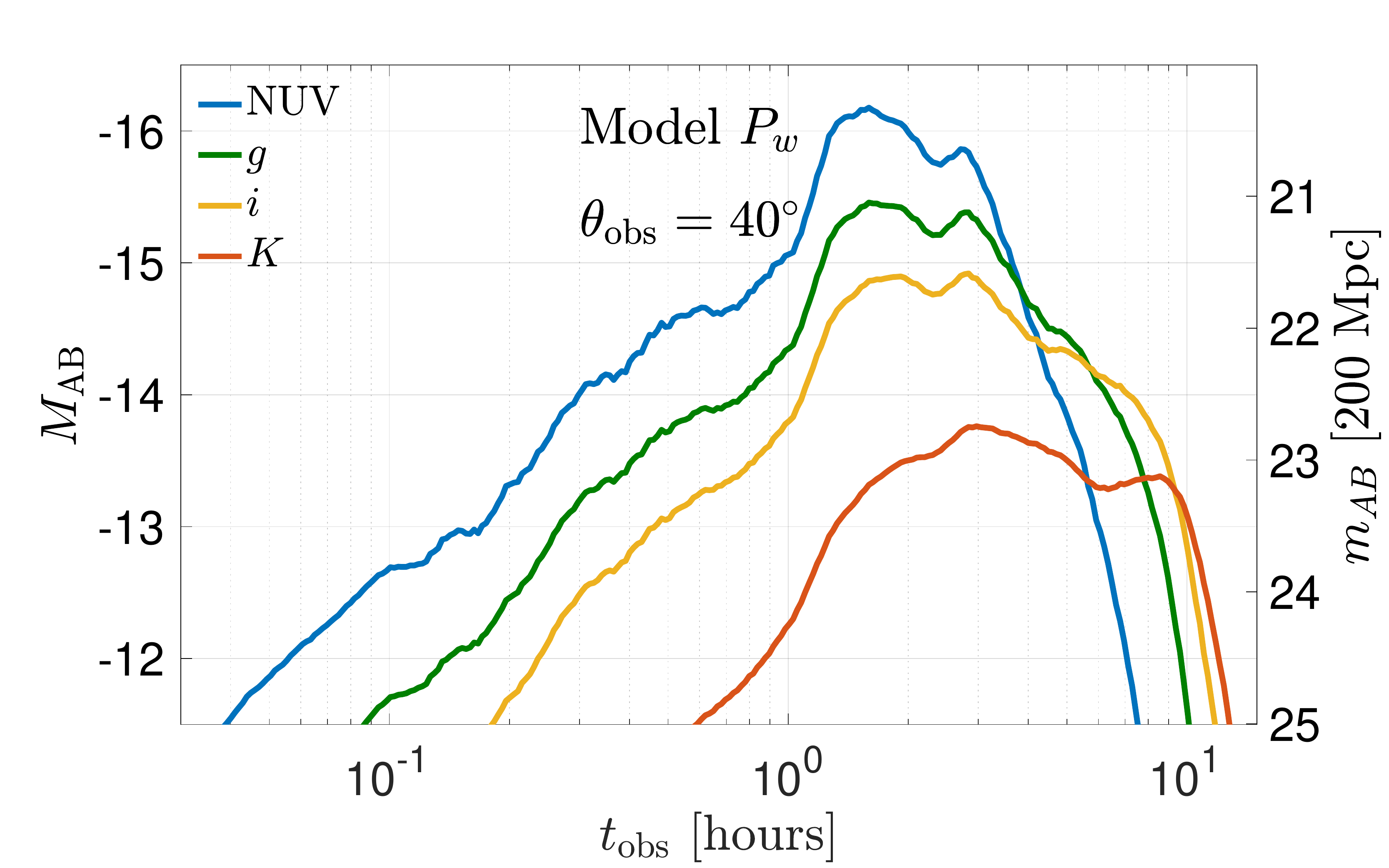}
    	\includegraphics[scale=0.26]{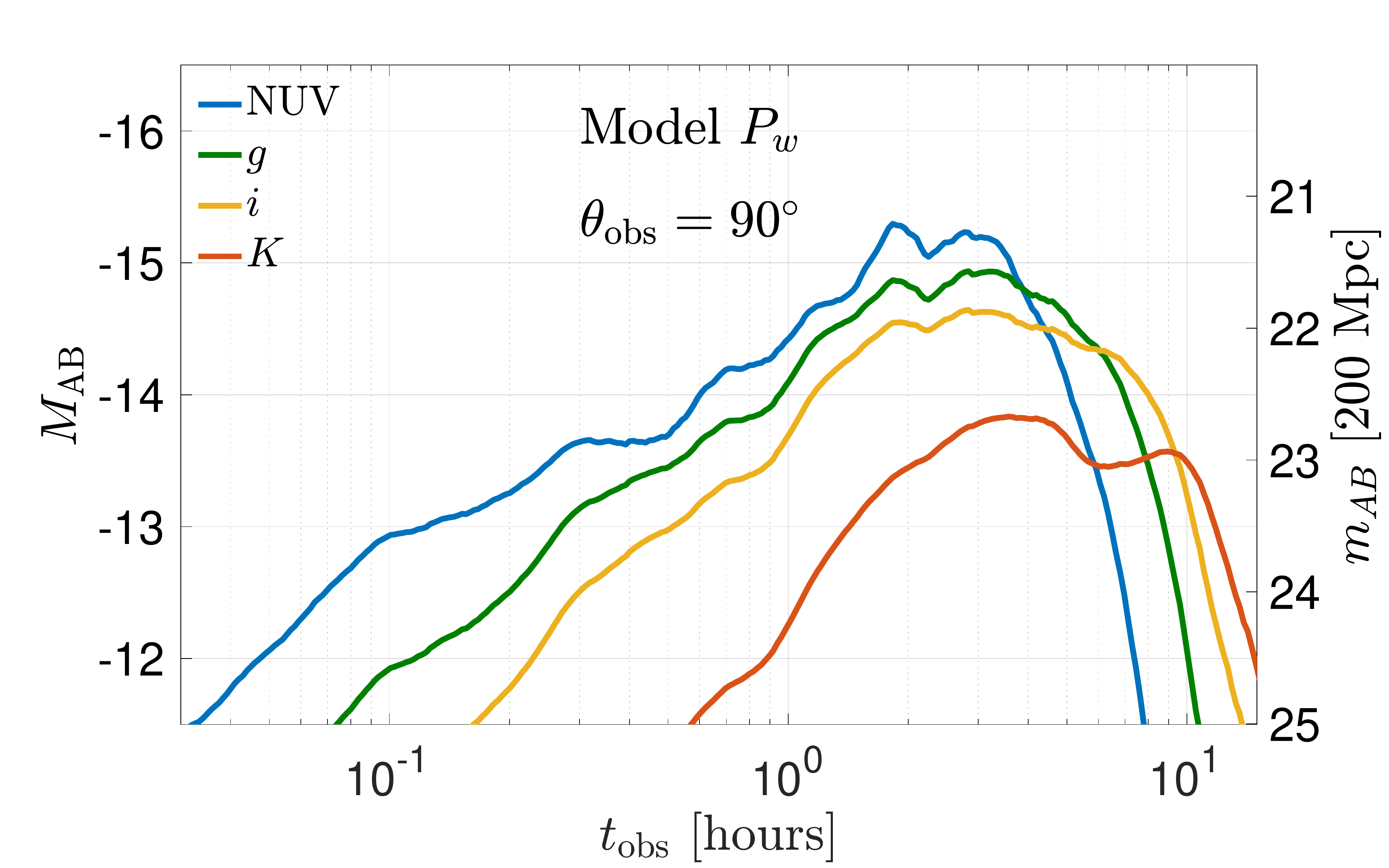}
    	\includegraphics[scale=0.26]{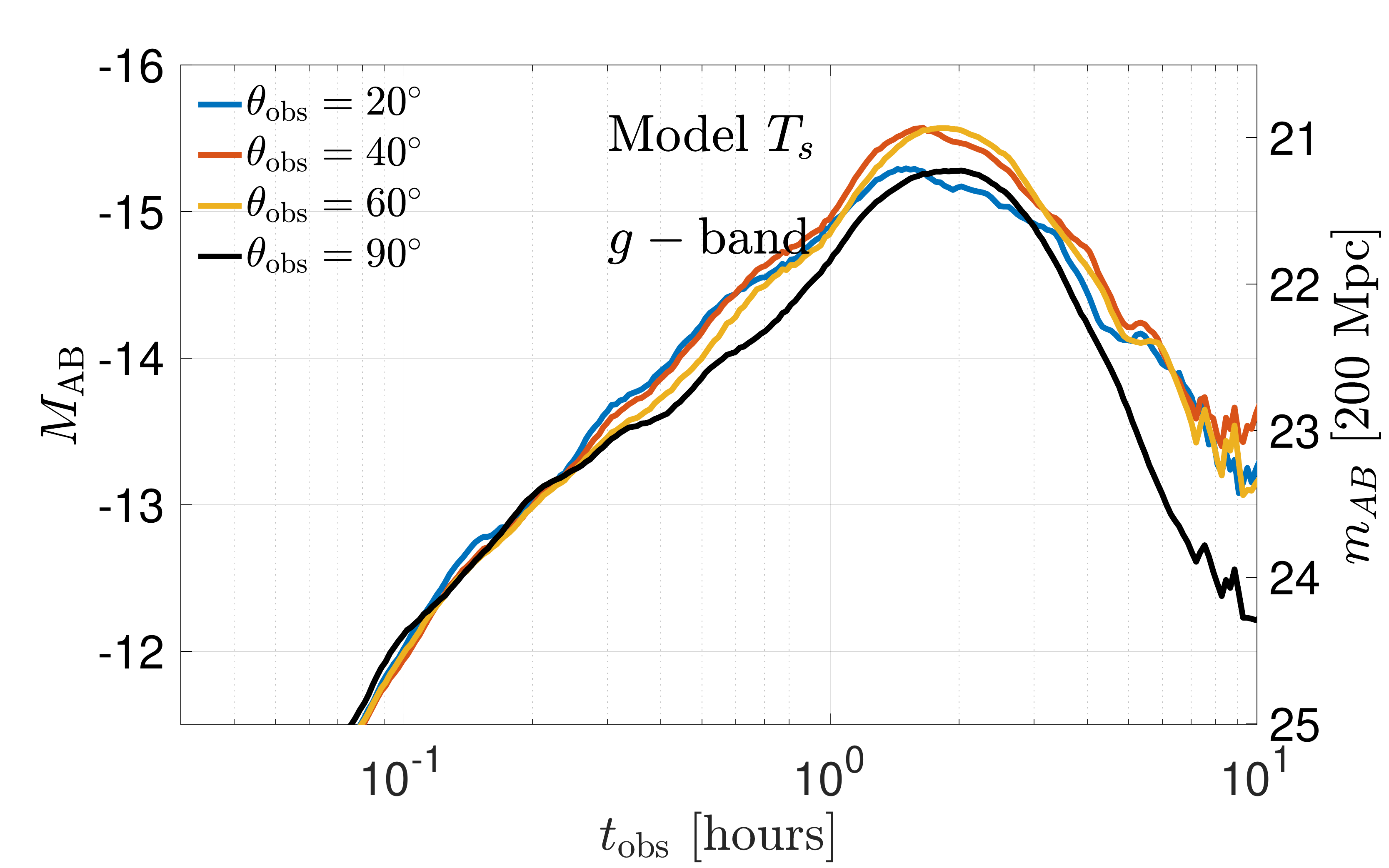}
    	\includegraphics[scale=0.26]{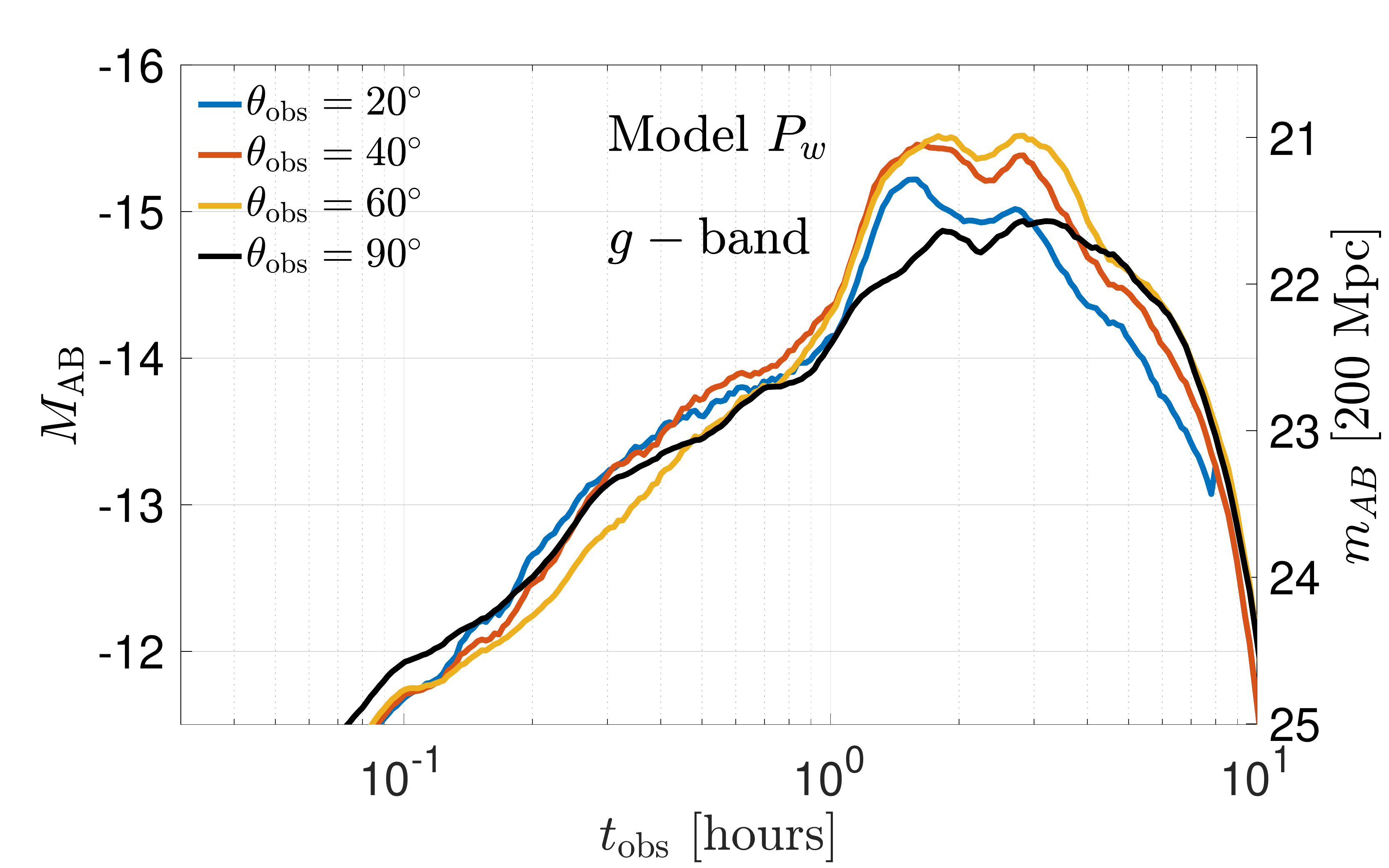}
     \caption{
     {\bf First row}: bolometric luminosity in different models at $ \thobs = 40^\circ $ (left) and at various viewing angles in model $ T_s $ (right).
     {\bf Second row}: $g$-band magnitude ($ \lambda = 475 $ nm) of different models at $ \thobs = 40^\circ $ (left) and $ \thobs = 90^\circ $ (right). Thin dashed lines represent the contribution of $r$-process heating outside of the trapping radius.
     {\bf Third row}: NUV ($ \lambda_{\rm NUV} = 250 $~nm), $g$ ($ \lambda_g = 475 $~nm), $i$ ($ \lambda_i = 760 $~nm), and $K$ ($ \lambda_K = 2160 $~nm) bands for model $ P_w $ at $ \thobs = 40^\circ $ (left) and $ \thobs = 90^\circ $ (right).
     {\bf Fourth row}: different angles in the $g$-band for model $ T_s $ (left) and $ P_w $ (right). The light curves are shown in absolute magnitude (left y-axis) and apparent magnitude at $ 200 $~Mpc (right y-axis).
     }
    	\label{fig:emission}
    \end{figure*}

Figure~\ref{fig:emission} depicts the emission profiles averaged over azimuthal angles\footnote{We find negligible variation in the observed light curves when considering different azimuthal angles. However, near the equator, where the emission is predominantly influenced by outflows from the disk, some asymmetry, such as in the tidal tail, leads to variations of up to half a magnitude. At $ \lvert\theta_{\rm obs}\rvert \lesssim 60^\circ $, the differences are up to $ \sim 0.2 $~mag.} as a function of the observer time. The first row shows the bolometric luminosity for different models and at various angles, while the lower panels present the light curves in absolute (left y-axis) and apparent at $ 200 $~Mpc (right y-axis) magnitudes in the $g$-band (second row), various bands for model $ P_w $ (third row), and at different angles in the $g$-band (fourth row). The light curve evolution can be described by two phases: optically thick and optically thin cocoon\footnote{Except for model $ H_0 $ where there is no cocoon, so only the first phase is present.}.

1. During the first minutes to an hour, the cocoon is still optically thick, and the emission is dominated by the front of the ejecta close to the equator. The strong initial fields in models $ T_s $ and $ P_s $ result in early ejection of a larger mass, leading to brighter emission. However, despite the high bolometric luminosity at early times (first row), the observed NUV/optical emission is still weak in all models (second row). This is because the color temperature is $ T \gtrsim 10^5~{\rm K} $, causing most of the radiation to exist as X-rays and far-UV photons. Furthermore, these photons cannot diffuse out to the observer due to the high opacity resulting from bound-free transitions \citep{Metzger2019}. This is not taken into account in our calculation; thus our calculation at early times is inapplicable. Similarly, the validity of the predicted early cocoon-powered X-ray emission proposed by \citet{Lazzati2017b,Lazzati2017a,Hamidani2022} is also called into question.

2. As the cocoon becomes optically thin, a hump appears in the light curve, indicating the release of a significant amount of internal energy leaking out as NUV/optical photons. A clear trend of earlier (Eq.~\eqref{eq:time}) and brighter emission at smaller viewing angles is apparent (top left panel), owing to lighter and more energetic gas close to the jet axis. As the temperature decreases, the NUV/optical light curve rises on a timescale of $ t_{\rm obs} \gtrsim 1 $~hour. Subsequently, the temperature stabilizes around $ T \approx 5\times 10^3~{\rm K} $ after a few hours, leading to a nearly monochromatic peak in the light curve (third row). As the emission is in the Rayleigh-Jeans regime, the light curves $ L_\nu \sim LT^{-3} $ dictate that a constant temperature results in a rapid decline of the light curve with the bolometric luminosity. During the decay in the light curve, the emitting gas is sub-relativistic, resulting in nearly isotropic emission (fourth row). Once the absolute magnitude drops below $ M_{\rm AB} \approx -13 $ at $ t_{\rm obs} \approx 10 $ hours, the main kilonova signal becomes the dominant source of emission \citep[see, e.g.,][]{Gompertz2023}. We do not incorporate the main kilonova component of the ejecta emission, as the slower parts of the ejecta have yet to reach homologous expansion at the time of the final simulation snapshots. Therefore, our calculations would no longer be valid beyond that point.

The second row of Fig.~\ref{fig:emission} demonstrates that the bright and long cooling emission (thick solid lines) outshines the negligible contribution of radioactive decay emission in the cocoon (thin dashed lines), i.e., $ L_r \ll L_c $, at all times. This finding differs from the results of \citet{Gottlieb2018a,Gottlieb2020c}, who found a shorter signal and a double-humped structure when $ \kappa \lesssim 1\,{\rm cm^2\,g^{-1}} $, attributing it to an early cooling signal followed by the kilonova emission from the cocoon. The disparities between our findings and theirs are attributed to two main factors. Firstly, the angular distribution of the electron fraction indicates that although the opacity is uncertain at these high temperatures, it seems to be relatively high, $ \kappa \gtrsim 5\,{\rm cm^2\,g^{-1}} $, resulting in a rather faint kilonova signal, akin to their models with $ \kappa = 10\,{\rm cm^2\,g^{-1}} $. Indeed, when $ \kappa = 1\,{\rm cm^2\,g^{-1}} $ (brown lines), the cooling emission peaks at half the time (see also Eq.~\eqref{eq:kappa}), however it still outshines the radioactive decay heating. Secondly, whereas they found the cocoon to contain only 10\% of the jet energy, our cocoons have a comparable energy to the jets and thus are an order-of-magnitude more massive. This is due to either more powerful outflows that unbind a larger amount of ejecta (e.g., model $ P_s $), or the late activation of a magnetically arrested disk (e.g., models $ T_s, P_s, P_w $) that prolongs the interaction between the jet and the ejecta. The cocoon-jet energy ratio depends on the amount of ejecta, among other things, which in turn depends on the binary properties. These factors could not be addressed in previous models, as the self-consistent modeling of jet launching was absent. An order-of-magnitude more massive cocoon explains why the signal is three times longer (Eq.~\eqref{eq:time}). We note that both of these aspects may be altered in other binary merger configurations, where the jet is likely less energetic \citep[see][]{Gottlieb2023}, or when neutrinos are incorporated into the post-merger simulation, as neutrino cooling influences both the electron fraction and the amount of unbound ejecta (see \S\ref{sec:discussion}).

Our results suggest that BH--NS mergers with the specific properties examined in this study are expected to be accompanied by an NUV/optical signal peaking at an apparent magnitude of $ \sim 21 \mbox{--} 22 $ in the UV/$g$-band, at the detection horizon of LVK run O4, $ D \approx 200 $~Mpc \citep{Abbott2020}. This magnitude makes the peak of closer events detectable by the Zwicky Transient Facility \citep[ZTF;][]{Graham2019,Dekany2020} at some viewing angles. Future telescopes such as the Rubin Observatory \citep{LSST2009} and ULTRASAT \citep{Ben-Ami2022,Shvartzvald2023} are projected to survey the sky down to single-visit magnitudes of $ \sim 25 $ and $ \sim 22 $, respectively. Rapid wide-field follow-up observations in the NUV and optical ranges, triggered by a GW detection from a BH--NS merger, could successfully detect these electromagnetic signals. Rubin Observatory will have the capacity to monitor the evolution of such signals within the LVK O4 detection horizon or detect parts of them at distances of $ D \gtrsim 1 $~Gpc.

\section{Discussion}\label{sec:discussion}

We conducted the first calculation of early NUV/optical emission originating from relativistic outflows that evolve self-consistently in compact binary mergers starting from the pre-merger phase. Notably, this is the first computation of such emission specifically in the context of BH--NS mergers, and the first to utilize M1 neutrino transport in large-scale jets. Our investigation involved tracking the evolution of an aligned BH--NS binary with a mass ratio of $ q = 2 $ and a BH dimensionless spin of $ a = 0.6 $, starting from the pre-merger and progressing to the homologous expansion. Subsequently, we advanced the system semi-analytically to the emission zone. By considering cooling envelope and $r$-process radioactive decay emission mechanisms, we calculated the light curves in the NUV/optical bands for different outflows and at various viewing angles.

Our findings indicate that in BH--NS mergers with low mass ratios and high BH spin, the presence of relativistic jets leads to a bright NUV/optical signal that lasts for a few hours. This signal peaks at an absolute magnitude of $M_{\rm AB} \approx -15 $, outshining any contribution from radioactive decay emission. Compared to previous studies of cocoon cooling emission that did not incorporate self-consistent jet launching, our results demonstrate a longer-lasting signal, owing to a more accurate estimation of the cocoon energy resulting from the interaction between the jet and ejecta.

The brightness of these signals may enable their detection by the ZTF out to $ \sim 100 $~Mpc. The upcoming Rubin Observatory, scheduled to begin observations during the later stages of O4, will be capable of effectively monitoring the entire evolution of such signals and detecting parts of them at distances beyond $D \gtrsim 1$ Gpc. Additionally, ULTRASAT, anticipated to launch in 2026, will cover the entire sky within minutes, enabling the detection of some of these signals out to $ \sim 200$~Mpc.

Detecting and interpreting the signals from these outflows poses a crucial question: what insights can we gain about the system? Differentiating between various types of outflows is a highly challenging task due to the complex and nonlinear nature of their evolution, as well as the inherent degeneracy in the system properties. The brightness of the signals is influenced by both the strength of the outflows, the viewing angle, and the optical depth of the outflow. We do not find any clear signature of the outflow structure or the initial magnetic field configuration in the light curves.

Another question to consider is whether we can distinguish BH--NS mergers from BNS mergers. Inferring the merger type based on the early emission from BH--NS mergers with a mass ratio of $ q = 2 $ may be unfeasible, as such systems might evolve in a similar way to BNS mergers if the BNS merger product collapses to a BH relatively fast.
However, BH--NS mergers can manifest in various forms with lower BH spins, higher mass ratios, and misaligned orbits. These variations lead to less mass remaining outside the BH innermost stable circular orbit after the merger, resulting in a reduced energy reservoir for jet launching and disk winds and, subsequently, a less energetic cocoon.

In the companion paper, \citet{Gottlieb2023}, we show that all jets exhibit either excessive luminosity or an extended launching process when compared to typical sGRBs. The reason behind this lies in the fact that, in order to achieve the expected luminosity of sGRBs, the jets must reach their maximum efficiency after a significant decrease in the accretion rate. However, this necessitates a longer duration for the jet launching process than what is typically observed in sGRBs. This poses a fundamental challenge to our current understanding of jet formation in binary mergers. A possible solution is that most sGRB jets in these binaries are launched from low pre-merger BH spin, as suggested by LVK observations \citep{Abbott2021}, or in higher mass ratio and spin-orbit misaligned binaries, as suggested by population synthesis models \citep{Belczynski2008}. In all such cases, the bound mass (disk) will be lower, such that the jet and cocoon (emission) will be weaker. Therefore, the aforementioned variations in the merger properties might be favored in order to fit sGRB observations. Thus, while a low mass ratio is similar to BNS mergers and ideal for cocoon emission, other BH--NS mergers might exhibit considerably earlier, shorter, and fainter emission, which could display distinct characteristics compared to BNS mergers.

An alternative method for deducing the nature of the system is to examine observational signatures that are exclusive to each merger type. For example, the role of the cocoon's kilonova, as previously proposed in \citet{Nakar2017,Gottlieb2018a,Gottlieb2020c,Hamidani2022}, remains uncertain, as we find here, and due to the considerable reliance on the poorly understood opacity of the hot cocoon, it might differ between BNS and BH--NS mergers. Additionally, if a fast tail of free neutrons is unique to BNS mergers, it could aid in identifying the merger type. This could be achieved through the detection of radioactive decay emission from free neutrons in the ejecta \citep{Metzger2015} or within the cocoon \citep{Gottlieb2020c}, although its emission likely overlaps with other mechanisms occurring on a similar timescale. Finally, when the cocoon breaks free from the ejecta, $ \gamma $-ray photons leak out through the breakout layer, generating a short-lived shock breakout signal, the characteristics of which strongly depend on the front ejecta mass and velocity \citep{Gottlieb2018b}. BH--NS mergers might exhibit a different front ejecta structure compared to BNS mergers, providing yet another electromagnetic counterpart that could aid in identifying the merger type. Calculation and comparison of the shock breakout emission in various merger configurations is a topic for future work.

To address the challenges mentioned above, we plan to
(i) conduct a comprehensive analysis spanning from pre-merger to post-merger stages of BH--NS systems with higher mass ratios and lower BH spins;
(ii) perform a self-consistent radiative transfer calculation of the emission, in order to properly resolve physical processes such as ionization of the material near the photosphere, reprocessing of photons, and non-radial photon diffusion \citep{Darbha2020,Darbha2021}; and
(iii) employ a complete neutrino transport scheme, which will enable us to properly account for the evolution of neutrinos within the system. We anticipate that neutrino cooling will suppress the early strong winds, leading to a reduction in wind emission and a decrease in the mass of the cocoon. As a result, the cocoon emission will also be attenuated. However, in \citet{Gottlieb2023}, we demonstrate that the winds are strongly influenced by magnetized outflows, indicating that they may not be significantly affected by neutrino cooling. The inclusion of neutrino transport will also allow us to self-consistently examine the evolution of the electron fraction within the cocoon. This will enable a more precise calculation of the resulting emission, improving the accuracy of our predictions.
                
\begin{acknowledgements}

OG is supported by a CIERA Postdoctoral Fellowship.
DI is supported by Future Investigators in NASA Earth and Space Science and Technology (FINESST) award No. 80NSSC21K1851. 
OG and AT acknowledge support by Fermi Cycle 14 Guest Investigator program 80NSSC22K0031.
JJ and AT acknowledge support by the NSF AST-2009884 and NASA 80NSSC21K1746 grants.
AT was also supported by NSF grants
AST-2107839, 
AST-1815304, 
AST-1911080, 
AST-2206471,
OAC-2031997, 
and NASA grant 80NSSC18K0565. 
RP acknowledges
support by NSF award AST-2006839.
Support for this work was also provided by the National Aeronautics and Space Administration through Chandra Award Number TM1-22005X issued by the Chandra X-ray Center, which is operated by the Smithsonian Astrophysical Observatory for and on behalf of the National Aeronautics Space Administration under contract NAS8-03060.  This research was facilitated by the Multimessenger Plasma Physics Center (MPPC), NSF grant PHY-2206607.
This research used resources of the Oak Ridge Leadership Computing Facility, which is a DOE Office of Science User Facility supported under contract DE-AC05-00OR22725. This research used resources of the National Energy Research Scientific Computing Center, a DOE Office of Science User Facility supported by the Office of Science of the U.S. Department of Energy under contract No. DE-AC02-05CH11231 using
NERSC award NP-ERCAP0020543 (allocation m2401). An award of computer time was provided by the ASCR Leadership Computing Challenge (ALCC), Innovative and Novel Computational Impact on Theory and Experiment (INCITE), and OLCF Director's Discretionary Allocation  programs under award PHY129. This research used resources of the National Energy Research Scientific Computing Center, a DOE Office of Science User Facility supported by the Office of Science of the U.S. Department of Energy under contract No. DE-AC02-05CH11231 using NERSC award ALCC-ERCAP0022634.
    
\end{acknowledgements}

\section*{Data Availability}

The data underlying this paper will be shared upon reasonable request to the corresponding author.
	
\bibliography{refs}

\begin{thebibliography}{}
\expandafter\ifx\csname natexlab\endcsname\relax\def\natexlab#1{#1}\fi
\providecommand{\url}[1]{\href{#1}{#1}}
\providecommand{\dodoi}[1]{doi:~\href{http://doi.org/#1}{\nolinkurl{#1}}}
\providecommand{\doeprint}[1]{\href{http://ascl.net/#1}{\nolinkurl{http://ascl.net/#1}}}
\providecommand{\doarXiv}[1]{\href{https://arxiv.org/abs/#1}{\nolinkurl{https://arxiv.org/abs/#1}}}

\bibitem[{{Abbott} {et~al.}(2020){Abbott}, {Abbott}, {Abbott}, {Abraham},
  {Acernese}, {Ackley}, {Adams}, {Adya}, {Affeldt}, {Agathos}, {Agatsuma},
  {Aggarwal}, {Aguiar}, {Aiello}, {Ain}, {Ajith}, {Akutsu}, {Allen}, {Allocca},
  {Aloy}, {Altin}, {Amato}, {Ananyeva}, {Anderson}, {Anderson}, {Ando},
  {Angelova}, {Antier}, {Appert}, {Arai}, {Arai}, {Arai}, {Araki}, {Araya},
  {Araya}, {Areeda}, {Ar{\`e}ne}, {Aritomi}, {Arnaud}, {Arun}, {Ascenzi},
  {Ashton}, {Aso}, {Aston}, {Astone}, {Aubin}, {Aufmuth}, {Aultoneal},
  {Austin}, {Avendano}, {Avila-Alvarez}, {Kagra Collaboration}, \& {VIRGO
  Collaboration}}]{Abbott2020}
{Abbott}, B.~P., {Abbott}, R., {Abbott}, T.~D., {et~al.} 2020, Living Reviews
  in Relativity, 23, 3, \dodoi{10.1007/s41114-020-00026-9}

\bibitem[{{Abbott} {et~al.}(2021){Abbott}, {Abbott}, {Abraham}, {Acernese},
  {Ackley}, {Adams}, {Adams}, {Adhikari}, {Adya}, {Affeldt}, {Agarwal},
  {Agathos}, {Agatsuma}, {Aggarwal}, {Aguiar}, {Aiello}, {Ain}, {Ajith},
  {Akutsu}, {Aleman}, {Allen}, {Allocca}, {Altin}, {Amato}, {Anand},
  {Ananyeva}, {Anderson}, {Anderson}, {Ando}, {Angelova}, {Ansoldi}, {Antelis},
  {Antier}, {Appert}, {Arai}, {Arai}, {Arai}, {Araki}, {Araya}, {Araya},
  {Areeda}, {Ar{\`e}ne}, {Aritomi}, {Arnaud}, {Aronson}, {Arun}, {Asada},
  {Asali}, {Ashton}, {Aso}, {Aston}, {Astone}, {Aubin}, {Aufmuth}, {Aultoneal},
  {Austin}, {Ligo Scientific Collaboration}, {VIRGO Collaboration}, \& {KAGRA
  Collaboration}}]{Abbott2021}
{Abbott}, R., {Abbott}, T.~D., {Abraham}, S., {et~al.} 2021, \apjl, 915, L5,
  \dodoi{10.3847/2041-8213/ac082e}

\bibitem[{{Arcavi}(2018)}]{Arcavi2018}
{Arcavi}, I. 2018, \apjl, 855, L23, \dodoi{10.3847/2041-8213/aab267}

\bibitem[{{Banerjee} {et~al.}(2023){Banerjee}, {Tanaka}, {Kato}, \&
  {Gaigalas}}]{Banerjee2023}
{Banerjee}, S., {Tanaka}, M., {Kato}, D., \& {Gaigalas}, G. 2023, arXiv
  e-prints, arXiv:2304.05810, \dodoi{10.48550/arXiv.2304.05810}

\bibitem[{{Banerjee} {et~al.}(2020){Banerjee}, {Tanaka}, {Kawaguchi}, {Kato},
  \& {Gaigalas}}]{Banerjee2020}
{Banerjee}, S., {Tanaka}, M., {Kawaguchi}, K., {Kato}, D., \& {Gaigalas}, G.
  2020, \apj, 901, 29, \dodoi{10.3847/1538-4357/abae61}

\bibitem[{{Belczynski} {et~al.}(2008){Belczynski}, {Taam}, {Rantsiou}, \& {van
  der Sluys}}]{Belczynski2008}
{Belczynski}, K., {Taam}, R.~E., {Rantsiou}, E., \& {van der Sluys}, M. 2008,
  \apj, 682, 474, \dodoi{10.1086/589609}

\bibitem[{{Ben-Ami} {et~al.}(2022){Ben-Ami}, {Shvartzvald}, {Waxman}, {Netzer},
  {Yaniv}, {Algranatti}, {Gal-Yam}, {Lapid}, {Ofek}, {Topaz}, {Arcavi}, {Asif},
  {Azaria}, {Bahalul}, {Barschke}, {Bastian-Querner}, {Berge}, {Berlea},
  {Buehler}, {Dittmar}, {Gelman}, {Giavitto}, {Guttman}, {Haces Crespo},
  {Heilbrunn}, {Kachergincky}, {Kaipachery}, {Kowalski}, {Kulkarni}, {Kumar},
  {K{\"u}sters}, {Liran}, {Miron-Salomon}, {Mor}, {Nir}, {Nitzan}, {Philipp},
  {Porelli}, {Sagiv}, {Schliwinski}, {Sprecher}, {De Simone}, {Stern}, {Stone},
  {Trakhtenbrot}, {Vasilev}, {Watson}, \& {Zappon}}]{Ben-Ami2022}
{Ben-Ami}, S., {Shvartzvald}, Y., {Waxman}, E., {et~al.} 2022, in Society of
  Photo-Optical Instrumentation Engineers (SPIE) Conference Series, Vol. 12181,
  Space Telescopes and Instrumentation 2022: Ultraviolet to Gamma Ray, ed.
  J.-W.~A. {den Herder}, S.~{Nikzad}, \& K.~{Nakazawa}, 1218105,
  \dodoi{10.1117/12.2629850}

\bibitem[{{Brege} {et~al.}(2018){Brege}, {Duez}, {Foucart}, {Deaton}, {Caro},
  {Hemberger}, {Kidder}, {O'Connor}, {Pfeiffer}, \& {Scheel}}]{Brege2018}
{Brege}, W., {Duez}, M.~D., {Foucart}, F., {et~al.} 2018, \prd, 98, 063009,
  \dodoi{10.1103/PhysRevD.98.063009}

\bibitem[{{Darbha} \& {Kasen}(2020)}]{Darbha2020}
{Darbha}, S., \& {Kasen}, D. 2020, \apj, 897, 150,
  \dodoi{10.3847/1538-4357/ab9a34}

\bibitem[{{Darbha} {et~al.}(2021){Darbha}, {Kasen}, {Foucart}, \&
  {Price}}]{Darbha2021}
{Darbha}, S., {Kasen}, D., {Foucart}, F., \& {Price}, D.~J. 2021, \apj, 915,
  69, \dodoi{10.3847/1538-4357/abff5d}

\bibitem[{{Dekany} {et~al.}(2020){Dekany}, {Smith}, {Riddle}, {Feeney},
  {Porter}, {Hale}, {Zolkower}, {Belicki}, {Kaye}, {Henning}, {Walters},
  {Cromer}, {Delacroix}, {Rodriguez}, {Reiley}, {Mao}, {Hover}, {Murphy},
  {Burruss}, {Baker}, {Kowalski}, {Reif}, {Mueller}, {Bellm}, {Graham}, \&
  {Kulkarni}}]{Dekany2020}
{Dekany}, R., {Smith}, R.~M., {Riddle}, R., {et~al.} 2020, \pasp, 132, 038001,
  \dodoi{10.1088/1538-3873/ab4ca2}

\bibitem[{{Dichiara} {et~al.}(2021){Dichiara}, {Becerra}, {Chase}, {Troja},
  {Lee}, {Watson}, {Butler}, {O'Connor}, {Pereyra}, {L{\'o}pez}, {Lien},
  {Gottlieb}, \& {Kutyrev}}]{Dichiara2021}
{Dichiara}, S., {Becerra}, R.~L., {Chase}, E.~A., {et~al.} 2021, \apjl, 923,
  L32, \dodoi{10.3847/2041-8213/ac4259}

\bibitem[{{Duez} {et~al.}(2010){Duez}, {Foucart}, {Kidder}, {Ott}, \&
  {Teukolsky}}]{Duez2010}
{Duez}, M.~D., {Foucart}, F., {Kidder}, L.~E., {Ott}, C.~D., \& {Teukolsky},
  S.~A. 2010, Classical and Quantum Gravity, 27, 114106,
  \dodoi{10.1088/0264-9381/27/11/114106}

\bibitem[{{Ekanger} {et~al.}(2023){Ekanger}, {Bhattacharya}, \&
  {Horiuchi}}]{Ekanger2023}
{Ekanger}, N., {Bhattacharya}, M., \& {Horiuchi}, S. 2023, arXiv e-prints,
  arXiv:2303.00765, \dodoi{10.48550/arXiv.2303.00765}

\bibitem[{{Etienne} {et~al.}(2008){Etienne}, {Faber}, {Liu}, {Shapiro},
  {Taniguchi}, \& {Baumgarte}}]{Etienne2008}
{Etienne}, Z.~B., {Faber}, J.~A., {Liu}, Y.~T., {et~al.} 2008, \prd, 77,
  084002, \dodoi{10.1103/PhysRevD.77.084002}

\bibitem[{{Etienne} {et~al.}(2012){Etienne}, {Paschalidis}, \&
  {Shapiro}}]{Etienne2012}
{Etienne}, Z.~B., {Paschalidis}, V., \& {Shapiro}, S.~L. 2012, \prd, 86,
  084026, \dodoi{10.1103/PhysRevD.86.084026}

\bibitem[{{Fern{\'a}ndez} {et~al.}(2017){Fern{\'a}ndez}, {Foucart}, {Kasen},
  {Lippuner}, {Desai}, \& {Roberts}}]{Fernandez2017}
{Fern{\'a}ndez}, R., {Foucart}, F., {Kasen}, D., {et~al.} 2017, Classical and
  Quantum Gravity, 34, 154001, \dodoi{10.1088/1361-6382/aa7a77}

\bibitem[{{Fern{\'a}ndez} {et~al.}(2015){Fern{\'a}ndez}, {Quataert}, {Schwab},
  {Kasen}, \& {Rosswog}}]{Fernandez2015}
{Fern{\'a}ndez}, R., {Quataert}, E., {Schwab}, J., {Kasen}, D., \& {Rosswog},
  S. 2015, \mnras, 449, 390, \dodoi{10.1093/mnras/stv238}

\bibitem[{{Foucart}(2012)}]{Foucart2012a}
{Foucart}, F. 2012, \prd, 86, 124007, \dodoi{10.1103/PhysRevD.86.124007}

\bibitem[{{Foucart} {et~al.}(2019){Foucart}, {Duez}, {Kidder}, {Nissanke},
  {Pfeiffer}, \& {Scheel}}]{Foucart2019}
{Foucart}, F., {Duez}, M.~D., {Kidder}, L.~E., {et~al.} 2019, \prd, 99, 103025,
  \dodoi{10.1103/PhysRevD.99.103025}

\bibitem[{{Foucart} {et~al.}(2012){Foucart}, {Duez}, {Kidder}, {Scheel},
  {Szilagyi}, \& {Teukolsky}}]{Foucart2012b}
---. 2012, \prd, 85, 044015, \dodoi{10.1103/PhysRevD.85.044015}

\bibitem[{{Foucart} {et~al.}(2011){Foucart}, {Duez}, {Kidder}, \&
  {Teukolsky}}]{Foucart2011}
{Foucart}, F., {Duez}, M.~D., {Kidder}, L.~E., \& {Teukolsky}, S.~A. 2011,
  \prd, 83, 024005, \dodoi{10.1103/PhysRevD.83.024005}

\bibitem[{{Foucart} {et~al.}(2018){Foucart}, {Hinderer}, \&
  {Nissanke}}]{Foucart2018}
{Foucart}, F., {Hinderer}, T., \& {Nissanke}, S. 2018, \prd, 98, 081501,
  \dodoi{10.1103/PhysRevD.98.081501}

\bibitem[{Foucart {et~al.}(2016)Foucart, O'Connor, Roberts, Kidder, Pfeiffer,
  \& Scheel}]{Foucart2016}
Foucart, F., O'Connor, E., Roberts, L., {et~al.} 2016, Physical Review D, 94,
  123016, \dodoi{10.1103/PhysRevD.94.123016}

\bibitem[{{Foucart} {et~al.}(2014){Foucart}, {Deaton}, {Duez}, {O'Connor},
  {Ott}, {Haas}, {Kidder}, {Pfeiffer}, {Scheel}, \& {Szilagyi}}]{Foucart2014}
{Foucart}, F., {Deaton}, M.~B., {Duez}, M.~D., {et~al.} 2014, \prd, 90, 024026,
  \dodoi{10.1103/PhysRevD.90.024026}

\bibitem[{{Foucart} {et~al.}(2015){Foucart}, {O'Connor}, {Roberts}, {Duez},
  {Haas}, {Kidder}, {Ott}, {Pfeiffer}, {Scheel}, \& {Szilagyi}}]{Foucart2015}
{Foucart}, F., {O'Connor}, E., {Roberts}, L., {et~al.} 2015, \prd, 91, 124021,
  \dodoi{10.1103/PhysRevD.91.124021}

\bibitem[{{Foucart} {et~al.}(2017){Foucart}, {Desai}, {Brege}, {Duez}, {Kasen},
  {Hemberger}, {Kidder}, {Pfeiffer}, \& {Scheel}}]{Foucart2017}
{Foucart}, F., {Desai}, D., {Brege}, W., {et~al.} 2017, Classical and Quantum
  Gravity, 34, 044002, \dodoi{10.1088/1361-6382/aa573b}

\bibitem[{{Fragione}(2021)}]{Fragione2021}
{Fragione}, G. 2021, \apjl, 923, L2, \dodoi{10.3847/2041-8213/ac3bcd}

\bibitem[{{Freiburghaus} {et~al.}(1999){Freiburghaus}, {Rosswog}, \&
  {Thielemann}}]{Freiburghaus1999}
{Freiburghaus}, C., {Rosswog}, S., \& {Thielemann}, F.~K. 1999, \apjl, 525,
  L121, \dodoi{10.1086/312343}

\bibitem[{Gompertz {et~al.}(2023)Gompertz, Nicholl, Smith, Harisankar, Pratten,
  Schmidt, \& Smith}]{Gompertz2023}
Gompertz, B.~P., Nicholl, M., Smith, J.~C., {et~al.} 2023, arXiv e-prints,
  arXiv:2305.07582, \dodoi{10.48550/arXiv.2305.07582}

\bibitem[{{Gottlieb} \& {Loeb}(2020)}]{Gottlieb2020c}
{Gottlieb}, O., \& {Loeb}, A. 2020, \mnras, 493, 1753,
  \dodoi{10.1093/mnras/staa363}

\bibitem[{{Gottlieb} {et~al.}(2018{\natexlab{a}}){Gottlieb}, {Nakar}, \&
  {Piran}}]{Gottlieb2018a}
{Gottlieb}, O., {Nakar}, E., \& {Piran}, T. 2018{\natexlab{a}}, \mnras, 473,
  576, \dodoi{10.1093/mnras/stx2357}

\bibitem[{{Gottlieb} {et~al.}(2018{\natexlab{b}}){Gottlieb}, {Nakar}, {Piran},
  \& {Hotokezaka}}]{Gottlieb2018b}
{Gottlieb}, O., {Nakar}, E., {Piran}, T., \& {Hotokezaka}, K.
  2018{\natexlab{b}}, \mnras, 479, 588, \dodoi{10.1093/mnras/sty1462}

\bibitem[{{Gottlieb} {et~al.}(2023){Gottlieb}, {Issa}, {Jacquemin-Ide},
  {Liska}, {Foucart}, {Tchekhovskoy}, {Metzger}, {Quataert}, {Perna}, {Kasen},
  {Duez}, {Kidder}, {Pfeiffer}, \& {Scheeli}}]{Gottlieb2023}
{Gottlieb}, O., {Issa}, D., {Jacquemin-Ide}, J., {et~al.} 2023, arXiv e-prints,
  arXiv:2306.14947.
\newblock \doarXiv{2306.14947}

\bibitem[{{Graham} {et~al.}(2019){Graham}, {Kulkarni}, {Bellm}, {Adams},
  {Barbarino}, {Blagorodnova}, {Bodewits}, {Bolin}, {Brady}, {Cenko}, {Chang},
  {Coughlin}, {De}, {Eadie}, {Farnham}, {Feindt}, {Franckowiak}, {Fremling},
  {Gezari}, {Ghosh}, {Goldstein}, {Golkhou}, {Goobar}, {Ho}, {Huppenkothen},
  {Ivezi{\'c}}, {Jones}, {Juric}, {Kaplan}, {Kasliwal}, {Kelley}, {Kupfer},
  {Lee}, {Lin}, {Lunnan}, {Mahabal}, {Miller}, {Ngeow}, {Nugent}, {Ofek},
  {Prince}, {Rauch}, {van Roestel}, {Schulze}, {Singer}, {Sollerman}, {Taddia},
  {Yan}, {Ye}, {Yu}, {Barlow}, {Bauer}, {Beck}, {Belicki}, {Biswas}, {Brinnel},
  {Brooke}, {Bue}, {Bulla}, {Burruss}, {Connolly}, {Cromer}, {Cunningham},
  {Dekany}, {Delacroix}, {Desai}, {Duev}, {Feeney}, {Flynn}, {Frederick},
  {Gal-Yam}, {Giomi}, {Groom}, {Hacopians}, {Hale}, {Helou}, {Henning},
  {Hover}, {Hillenbrand}, {Howell}, {Hung}, {Imel}, {Ip}, {Jackson}, {Kaspi},
  {Kaye}, {Kowalski}, {Kramer}, {Kuhn}, {Landry}, {Laher}, {Mao}, {Masci},
  {Monkewitz}, {Murphy}, {Nordin}, {Patterson}, {Penprase}, {Porter},
  {Rebbapragada}, {Reiley}, {Riddle}, {Rigault}, {Rodriguez}, {Rusholme}, {van
  Santen}, {Shupe}, {Smith}, {Soumagnac}, {Stein}, {Surace}, {Szkody}, {Terek},
  {Van Sistine}, {van Velzen}, {Vestrand}, {Walters}, {Ward}, {Zhang}, \&
  {Zolkower}}]{Graham2019}
{Graham}, M.~J., {Kulkarni}, S.~R., {Bellm}, E.~C., {et~al.} 2019, \pasp, 131,
  078001, \dodoi{10.1088/1538-3873/ab006c}

\bibitem[{{Hamidani} \& {Ioka}(2022)}]{Hamidani2022}
{Hamidani}, H., \& {Ioka}, K. 2022, arXiv e-prints, arXiv:2210.02255,
  \dodoi{10.48550/arXiv.2210.02255}

\bibitem[{{Hayashi} {et~al.}(2022{\natexlab{a}}){Hayashi}, {Fujibayashi},
  {Kiuchi}, {Kyutoku}, {Sekiguchi}, \& {Shibata}}]{Hayashi2022a}
{Hayashi}, K., {Fujibayashi}, S., {Kiuchi}, K., {et~al.} 2022{\natexlab{a}},
  \prd, 106, 023008, \dodoi{10.1103/PhysRevD.106.023008}

\bibitem[{{Hayashi} {et~al.}(2021){Hayashi}, {Kawaguchi}, {Kiuchi}, {Kyutoku},
  \& {Shibata}}]{Hayashi2021}
{Hayashi}, K., {Kawaguchi}, K., {Kiuchi}, K., {Kyutoku}, K., \& {Shibata}, M.
  2021, \prd, 103, 043007, \dodoi{10.1103/PhysRevD.103.043007}

\bibitem[{{Hayashi} {et~al.}(2022{\natexlab{b}}){Hayashi}, {Kiuchi}, {Kyutoku},
  {Sekiguchi}, \& {Shibata}}]{Hayashi2022b}
{Hayashi}, K., {Kiuchi}, K., {Kyutoku}, K., {Sekiguchi}, Y., \& {Shibata}, M.
  2022{\natexlab{b}}, arXiv e-prints, arXiv:2211.07158,
  \dodoi{10.48550/arXiv.2211.07158}

\bibitem[{{Hotokezaka} {et~al.}(2016){Hotokezaka}, {Wanajo}, {Tanaka}, {Bamba},
  {Terada}, \& {Piran}}]{Hotokezaka2016b}
{Hotokezaka}, K., {Wanajo}, S., {Tanaka}, M., {et~al.} 2016, \mnras, 459, 35,
  \dodoi{10.1093/mnras/stw404}

\bibitem[{{Janka} {et~al.}(1999){Janka}, {Eberl}, {Ruffert}, \&
  {Fryer}}]{Janka1999}
{Janka}, H.~T., {Eberl}, T., {Ruffert}, M., \& {Fryer}, C.~L. 1999, \apjl, 527,
  L39, \dodoi{10.1086/312397}

\bibitem[{{Kasen} {et~al.}(2013){Kasen}, {Badnell}, \& {Barnes}}]{Kasen2013}
{Kasen}, D., {Badnell}, N.~R., \& {Barnes}, J. 2013, \apj, 774, 25,
  \dodoi{10.1088/0004-637X/774/1/25}

\bibitem[{{Kasliwal} {et~al.}(2017){Kasliwal}, {Nakar}, {Singer}, {Kaplan},
  {Cook}, {Van Sistine}, {Lau}, {Fremling}, {Gottlieb}, {Jencson}, {Adams},
  {Feindt}, {Hotokezaka}, {Ghosh}, {Perley}, {Yu}, {Piran}, {Allison},
  {Anupama}, {Balasubramanian}, {Bannister}, {Bally}, {Barnes}, {Barway},
  {Bellm}, {Bhalerao}, {Bhattacharya}, {Blagorodnova}, {Bloom}, {Brady},
  {Cannella}, {Chatterjee}, {Cenko}, {Cobb}, {Copperwheat}, {Corsi}, {De},
  {Dobie}, {Emery}, {Evans}, {Fox}, {Frail}, {Frohmaier}, {Goobar}, {Hallinan},
  {Harrison}, {Helou}, {Hinderer}, {Ho}, {Horesh}, {Ip}, {Itoh}, {Kasen},
  {Kim}, {Kuin}, {Kupfer}, {Lynch}, {Madsen}, {Mazzali}, {Miller}, {Mooley},
  {Murphy}, {Ngeow}, {Nichols}, {Nissanke}, {Nugent}, {Ofek}, {Qi}, {Quimby},
  {Rosswog}, {Rusu}, {Sadler}, {Schmidt}, {Sollerman}, {Steele}, {Williamson},
  {Xu}, {Yan}, {Yatsu}, {Zhang}, \& {Zhao}}]{Kasliwal2017}
{Kasliwal}, M.~M., {Nakar}, E., {Singer}, L.~P., {et~al.} 2017, Science, 358,
  1559, \dodoi{10.1126/science.aap9455}

\bibitem[{{Kawaguchi} {et~al.}(2015){Kawaguchi}, {Kyutoku}, {Nakano}, {Okawa},
  {Shibata}, \& {Taniguchi}}]{Kawaguchi2015}
{Kawaguchi}, K., {Kyutoku}, K., {Nakano}, H., {et~al.} 2015, \prd, 92, 024014,
  \dodoi{10.1103/PhysRevD.92.024014}

\bibitem[{{Kawaguchi} {et~al.}(2016){Kawaguchi}, {Kyutoku}, {Shibata}, \&
  {Tanaka}}]{Kawaguchi2016}
{Kawaguchi}, K., {Kyutoku}, K., {Shibata}, M., \& {Tanaka}, M. 2016, \apj, 825,
  52, \dodoi{10.3847/0004-637X/825/1/52}

\bibitem[{{Kiuchi} {et~al.}(2015){Kiuchi}, {Cerd{\'a}-Dur{\'a}n}, {Kyutoku},
  {Sekiguchi}, \& {Shibata}}]{Kiuchi2015}
{Kiuchi}, K., {Cerd{\'a}-Dur{\'a}n}, P., {Kyutoku}, K., {Sekiguchi}, Y., \&
  {Shibata}, M. 2015, \prd, 92, 124034, \dodoi{10.1103/PhysRevD.92.124034}

\bibitem[{{Korobkin} {et~al.}(2012){Korobkin}, {Rosswog}, {Arcones}, \&
  {Winteler}}]{Korobkin2012}
{Korobkin}, O., {Rosswog}, S., {Arcones}, A., \& {Winteler}, C. 2012, \mnras,
  426, 1940, \dodoi{10.1111/j.1365-2966.2012.21859.x}

\bibitem[{{Kyutoku} {et~al.}(2015){Kyutoku}, {Ioka}, {Okawa}, {Shibata}, \&
  {Taniguchi}}]{Kyutoku2015}
{Kyutoku}, K., {Ioka}, K., {Okawa}, H., {Shibata}, M., \& {Taniguchi}, K. 2015,
  \prd, 92, 044028, \dodoi{10.1103/PhysRevD.92.044028}

\bibitem[{{Kyutoku} {et~al.}(2013){Kyutoku}, {Ioka}, \&
  {Shibata}}]{Kyutoku2013}
{Kyutoku}, K., {Ioka}, K., \& {Shibata}, M. 2013, \prd, 88, 041503,
  \dodoi{10.1103/PhysRevD.88.041503}

\bibitem[{{Kyutoku} {et~al.}(2018){Kyutoku}, {Kiuchi}, {Sekiguchi}, {Shibata},
  \& {Taniguchi}}]{Kyutoku2018}
{Kyutoku}, K., {Kiuchi}, K., {Sekiguchi}, Y., {Shibata}, M., \& {Taniguchi}, K.
  2018, \prd, 97, 023009, \dodoi{10.1103/PhysRevD.97.023009}

\bibitem[{{Kyutoku} {et~al.}(2011){Kyutoku}, {Okawa}, {Shibata}, \&
  {Taniguchi}}]{Kyutoku2011}
{Kyutoku}, K., {Okawa}, H., {Shibata}, M., \& {Taniguchi}, K. 2011, \prd, 84,
  064018, \dodoi{10.1103/PhysRevD.84.064018}

\bibitem[{{Lazzati} {et~al.}(2017{\natexlab{a}}){Lazzati}, {Deich}, {Morsony},
  \& {Workman}}]{Lazzati2017b}
{Lazzati}, D., {Deich}, A., {Morsony}, B.~J., \& {Workman}, J.~C.
  2017{\natexlab{a}}, \mnras, 471, 1652, \dodoi{10.1093/mnras/stx1683}

\bibitem[{{Lazzati} {et~al.}(2017{\natexlab{b}}){Lazzati},
  {L{\'o}pez-C{\'a}mara}, {Cantiello}, {Morsony}, {Perna}, \&
  {Workman}}]{Lazzati2017a}
{Lazzati}, D., {L{\'o}pez-C{\'a}mara}, D., {Cantiello}, M., {et~al.}
  2017{\natexlab{b}}, \apjl, 848, L6, \dodoi{10.3847/2041-8213/aa8f3d}

\bibitem[{{Lazzati} {et~al.}(2018){Lazzati}, {Perna}, {Morsony},
  {Lopez-Camara}, {Cantiello}, {Ciolfi}, {Giacomazzo}, \&
  {Workman}}]{Lazzati2018}
{Lazzati}, D., {Perna}, R., {Morsony}, B.~J., {et~al.} 2018, \prl, 120, 241103,
  \dodoi{10.1103/PhysRevLett.120.241103}

\bibitem[{{Liska} {et~al.}(2022){Liska}, {Chatterjee}, {Issa}, {Yoon}, {Kaaz},
  {Tchekhovskoy}, {van Eijnatten}, {Musoke}, {Hesp}, {Rohoza}, {Markoff},
  {Ingram}, \& {van der Klis}}]{Liska2022}
{Liska}, M.~T.~P., {Chatterjee}, K., {Issa}, D., {et~al.} 2022, \apjs, 263, 26,
  \dodoi{10.3847/1538-4365/ac9966}

\bibitem[{{LSST Science Collaboration} {et~al.}(2009){LSST Science
  Collaboration}, {Abell}, {Allison}, {Anderson}, {Andrew}, {Angel}, {Armus},
  {Arnett}, {Asztalos}, {Axelrod}, {Bailey}, {Ballantyne}, {Bankert},
  {Barkhouse}, {Barr}, {Barrientos}, {Barth}, {Bartlett}, {Becker}, {Becla},
  {Beers}, {Bernstein}, {Biswas}, {Blanton}, {Bloom}, {Bochanski}, {Boeshaar},
  {Borne}, {Bradac}, {Brandt}, {Bridge}, {Brown}, {Brunner}, {Bullock},
  {Burgasser}, {Burge}, {Burke}, {Cargile}, {Chandrasekharan}, {Chartas},
  {Chesley}, {Chu}, {Cinabro}, {Claire}, {Claver}, {Clowe}, {Connolly}, {Cook},
  {Cooke}, {Cooray}, {Covey}, {Culliton}, {de Jong}, {de Vries}, {Debattista},
  {Delgado}, {Dell'Antonio}, {Dhital}, {Di Stefano}, {Dickinson}, {Dilday},
  {Djorgovski}, {Dobler}, {Donalek}, {Dubois-Felsmann}, {Durech},
  {Eliasdottir}, {Eracleous}, {Eyer}, {Falco}, {Fan}, {Fassnacht}, {Ferguson},
  {Fernandez}, {Fields}, {Finkbeiner}, {Figueroa}, {Fox}, {Francke}, {Frank},
  {Frieman}, {Fromenteau}, {Furqan}, {Galaz}, {Gal-Yam}, {Garnavich},
  {Gawiser}, {Geary}, {Gee}, {Gibson}, {Gilmore}, {Grace}, {Green}, {Gressler},
  {Grillmair}, {Habib}, {Haggerty}, {Hamuy}, {Harris}, {Hawley}, {Heavens},
  {Hebb}, {Henry}, {Hileman}, {Hilton}, {Hoadley}, {Holberg}, {Holman},
  {Howell}, {Infante}, {Ivezic}, {Jacoby}, {Jain}, {R}, {Jedicke}, {Jee},
  {Garrett Jernigan}, {Jha}, {Johnston}, {Jones}, {Juric}, {Kaasalainen},
  {Styliani}, {Kafka}, {Kahn}, {Kaib}, {Kalirai}, {Kantor}, {Kasliwal},
  {Keeton}, {Kessler}, {Knezevic}, {Kowalski}, {Krabbendam}, {Krughoff},
  {Kulkarni}, {Kuhlman}, {Lacy}, {Lepine}, {Liang}, {Lien}, {Lira}, {Long},
  {Lorenz}, {Lotz}, {Lupton}, {Lutz}, {Macri}, {Mahabal}, {Mandelbaum},
  {Marshall}, {May}, {McGehee}, {Meadows}, {Meert}, {Milani}, {Miller},
  {Miller}, {Mills}, {Minniti}, {Monet}, {Mukadam}, {Nakar}, {Neill}, {Newman},
  {Nikolaev}, {Nordby}, {O'Connor}, {Oguri}, {Oliver}, {Olivier}, {Olsen},
  {Olsen}, {Olszewski}, {Oluseyi}, {Padilla}, {Parker}, {Pepper}, {Peterson},
  {Petry}, {Pinto}, {Pizagno}, {Popescu}, {Prsa}, {Radcka}, {Raddick},
  {Rasmussen}, {Rau}, {Rho}, {Rhoads}, {Richards}, {Ridgway}, {Robertson},
  {Roskar}, {Saha}, {Sarajedini}, {Scannapieco}, {Schalk}, {Schindler},
  {Schmidt}, {Schmidt}, {Schneider}, {Schumacher}, {Scranton}, {Sebag},
  {Seppala}, {Shemmer}, {Simon}, {Sivertz}, {Smith}, {Allyn Smith}, {Smith},
  {Spitz}, {Stanford}, {Stassun}, {Strader}, {Strauss}, {Stubbs}, {Sweeney},
  {Szalay}, {Szkody}, {Takada}, {Thorman}, {Trilling}, {Trimble}, {Tyson}, {Van
  Berg}, {Vanden Berk}, {VanderPlas}, {Verde}, {Vrsnak}, {Walkowicz},
  {Wandelt}, {Wang}, {Wang}, {Warner}, {Wechsler}, {West}, {Wiecha},
  {Williams}, {Willman}, {Wittman}, {Wolff}, {Wood-Vasey}, {Wozniak}, {Young},
  {Zentner}, \& {Zhan}}]{LSST2009}
{LSST Science Collaboration}, {Abell}, P.~A., {Allison}, J., {et~al.} 2009,
  arXiv e-prints, arXiv:0912.0201, \dodoi{10.48550/arXiv.0912.0201}

\bibitem[{{Margutti} \& {Chornock}(2021)}]{Margutti2021}
{Margutti}, R., \& {Chornock}, R. 2021, \araa, 59,
  \dodoi{10.1146/annurev-astro-112420-030742}

\bibitem[{{Metzger}(2019)}]{Metzger2019}
{Metzger}, B.~D. 2019, Living Reviews in Relativity, 23, 1,
  \dodoi{10.1007/s41114-019-0024-0}

\bibitem[{{Metzger} {et~al.}(2010{\natexlab{a}}){Metzger}, {Arcones},
  {Quataert}, \& {Mart{\'\i}nez-Pinedo}}]{Metzger2010b}
{Metzger}, B.~D., {Arcones}, A., {Quataert}, E., \& {Mart{\'\i}nez-Pinedo}, G.
  2010{\natexlab{a}}, \mnras, 402, 2771,
  \dodoi{10.1111/j.1365-2966.2009.16107.x}

\bibitem[{{Metzger} {et~al.}(2015){Metzger}, {Bauswein}, {Goriely}, \&
  {Kasen}}]{Metzger2015}
{Metzger}, B.~D., {Bauswein}, A., {Goriely}, S., \& {Kasen}, D. 2015, \mnras,
  446, 1115, \dodoi{10.1093/mnras/stu2225}

\bibitem[{{Metzger} {et~al.}(2010{\natexlab{b}}){Metzger},
  {Mart{\'\i}nez-Pinedo}, {Darbha}, {Quataert}, {Arcones}, {Kasen}, {Thomas},
  {Nugent}, {Panov}, \& {Zinner}}]{Metzger2010a}
{Metzger}, B.~D., {Mart{\'\i}nez-Pinedo}, G., {Darbha}, S., {et~al.}
  2010{\natexlab{b}}, \mnras, 406, 2650,
  \dodoi{10.1111/j.1365-2966.2010.16864.x}

\bibitem[{{Mochkovitch} {et~al.}(1993){Mochkovitch}, {Hernanz}, {Isern}, \&
  {Martin}}]{Mochkovitch1993}
{Mochkovitch}, R., {Hernanz}, M., {Isern}, J., \& {Martin}, X. 1993, \nat, 361,
  236, \dodoi{10.1038/361236a0}

\bibitem[{{Mooley} {et~al.}(2018{\natexlab{a}}){Mooley}, {Nakar}, {Hotokezaka},
  {Hallinan}, {Corsi}, {Frail}, {Horesh}, {Murphy}, {Lenc}, {Kaplan}, {de},
  {Dobie}, {Chandra}, {Deller}, {Gottlieb}, {Kasliwal}, {Kulkarni}, {Myers},
  {Nissanke}, {Piran}, {Lynch}, {Bhalerao}, {Bourke}, {Bannister}, \&
  {Singer}}]{Mooley2018a}
{Mooley}, K.~P., {Nakar}, E., {Hotokezaka}, K., {et~al.} 2018{\natexlab{a}},
  \nat, 554, 207, \dodoi{10.1038/nature25452}

\bibitem[{{Mooley} {et~al.}(2018{\natexlab{b}}){Mooley}, {Deller}, {Gottlieb},
  {Nakar}, {Hallinan}, {Bourke}, {Frail}, {Horesh}, {Corsi}, \&
  {Hotokezaka}}]{Mooley2018b}
{Mooley}, K.~P., {Deller}, A.~T., {Gottlieb}, O., {et~al.} 2018{\natexlab{b}},
  \nat, 561, 355, \dodoi{10.1038/s41586-018-0486-3}

\bibitem[{{Most} {et~al.}(2021){Most}, {Papenfort}, {Tootle}, \&
  {Rezzolla}}]{Most2021}
{Most}, E.~R., {Papenfort}, L.~J., {Tootle}, S.~D., \& {Rezzolla}, L. 2021,
  \mnras, 506, 3511, \dodoi{10.1093/mnras/stab1824}

\bibitem[{{Nakar}(2020)}]{Nakar2019}
{Nakar}, E. 2020, \physrep, 886, 1, \dodoi{10.1016/j.physrep.2020.08.008}

\bibitem[{{Nakar} \& {Piran}(2017)}]{Nakar2017}
{Nakar}, E., \& {Piran}, T. 2017, \apj, 834, 28,
  \dodoi{10.3847/1538-4357/834/1/28}

\bibitem[{{O'Connor}(2015)}]{OConnor_Nulib}
{O'Connor}, E. 2015, \apjs, 219, 24, \dodoi{10.1088/0067-0049/219/2/24}

\bibitem[{{Paczynski}(1991)}]{Paczynski1991}
{Paczynski}, B. 1991, \actaa, 41, 257

\bibitem[{{Paschalidis} {et~al.}(2015){Paschalidis}, {Ruiz}, \&
  {Shapiro}}]{Paschalidis2015}
{Paschalidis}, V., {Ruiz}, M., \& {Shapiro}, S.~L. 2015, \apjl, 806, L14,
  \dodoi{10.1088/2041-8205/806/1/L14}

\bibitem[{{Perna} {et~al.}(2016){Perna}, {Lazzati}, \&
  {Giacomazzo}}]{Perna2016}
{Perna}, R., {Lazzati}, D., \& {Giacomazzo}, B. 2016, \apjl, 821, L18,
  \dodoi{10.3847/2041-8205/821/1/L18}

\bibitem[{{Piro} \& {Kollmeier}(2018)}]{Piro2018}
{Piro}, A.~L., \& {Kollmeier}, J.~A. 2018, \apj, 855, 103,
  \dodoi{10.3847/1538-4357/aaaab3}

\bibitem[{{Rantsiou} {et~al.}(2008){Rantsiou}, {Kobayashi}, {Laguna}, \&
  {Rasio}}]{Rantsiou2008}
{Rantsiou}, E., {Kobayashi}, S., {Laguna}, P., \& {Rasio}, F.~A. 2008, \apj,
  680, 1326, \dodoi{10.1086/587858}

\bibitem[{{Rastinejad} {et~al.}(2022){Rastinejad}, {Gompertz}, {Levan}, {Fong},
  {Nicholl}, {Lamb}, {Malesani}, {Nugent}, {Oates}, {Tanvir}, {de Ugarte
  Postigo}, {Kilpatrick}, {Moore}, {Metzger}, {Ravasio}, {Rossi}, {Schroeder},
  {Jencson}, {Sand}, {Smith}, {Ag{\"u}{\'\i} Fern{\'a}ndez}, {Berger},
  {Blanchard}, {Chornock}, {Cobb}, {De Pasquale}, {Fynbo}, {Izzo}, {Kann},
  {Laskar}, {Marini}, {Paterson}, {Rouco Escorial}, {Sears}, \&
  {Th{\"o}ne}}]{Rastinejad2022}
{Rastinejad}, J.~C., {Gompertz}, B.~P., {Levan}, A.~J., {et~al.} 2022, arXiv
  e-prints, arXiv:2204.10864.
\newblock \doarXiv{2204.10864}

\bibitem[{{Rosswog}(2005)}]{Rosswog2005}
{Rosswog}, S. 2005, \apj, 634, 1202, \dodoi{10.1086/497062}

\bibitem[{{Ruiz} {et~al.}(2018){Ruiz}, {Shapiro}, \& {Tsokaros}}]{Ruiz2018}
{Ruiz}, M., {Shapiro}, S.~L., \& {Tsokaros}, A. 2018, \prd, 98, 123017,
  \dodoi{10.1103/PhysRevD.98.123017}

\bibitem[{{Shibata} \& {Taniguchi}(2008)}]{Shibata2008}
{Shibata}, M., \& {Taniguchi}, K. 2008, \prd, 77, 084015,
  \dodoi{10.1103/PhysRevD.77.084015}

\bibitem[{{Shibata} \& {Taniguchi}(2011)}]{Shibata2011}
---. 2011, Living Reviews in Relativity, 14, 6, \dodoi{10.12942/lrr-2011-6}

\bibitem[{{Shibata} \& {Ury{\={u}}}(2006)}]{Shibata2006}
{Shibata}, M., \& {Ury{\={u}}}, K. 2006, \prd, 74, 121503,
  \dodoi{10.1103/PhysRevD.74.121503}

\bibitem[{{Shibata} \& {Ury{\={u}}}(2007)}]{Shibata2007}
---. 2007, Classical and Quantum Gravity, 24, S125,
  \dodoi{10.1088/0264-9381/24/12/S09}

\bibitem[{{Shvartzvald} {et~al.}(2023){Shvartzvald}, {Waxman}, {Gal-Yam},
  {Ofek}, {Ben-Ami}, {Berge}, {Kowalski}, {B{\"u}hler}, {Worm}, {Rhoads},
  {Arcavi}, {Maoz}, {Polishook}, {Stone}, {Trakhtenbrot}, {Ackermann},
  {Aharonson}, {Birnholtz}, {Chelouche}, {Guetta}, {Hallakoun}, {Horesh},
  {Kushnir}, {Mazeh}, {Nordin}, {Ofir}, {Ohm}, {Parsons}, {Pe'er}, {Perets},
  {Perdelwitz}, {Poznanski}, {Sadeh}, {Sagiv}, {Shahaf}, {Soumagnac}, {Tal-Or},
  {Van Santen}, {Zackay}, {Guttman}, {Rekhi}, {Townsend}, {Weinstein}, \&
  {Wold}}]{Shvartzvald2023}
{Shvartzvald}, Y., {Waxman}, E., {Gal-Yam}, A., {et~al.} 2023, arXiv e-prints,
  arXiv:2304.14482, \dodoi{10.48550/arXiv.2304.14482}

\bibitem[{{SpEC collaboration}(2023)}]{spec}
{SpEC collaboration}. 2023.
\newblock \url{https://www.black-holes.org/code/SpEC.html}

\bibitem[{{Surman} {et~al.}(2008){Surman}, {McLaughlin}, {Ruffert}, {Janka}, \&
  {Hix}}]{Surman2008}
{Surman}, R., {McLaughlin}, G.~C., {Ruffert}, M., {Janka}, H.~T., \& {Hix},
  W.~R. 2008, \apjl, 679, L117, \dodoi{10.1086/589507}

\bibitem[{{Tanaka} {et~al.}(2014){Tanaka}, {Hotokezaka}, {Kyutoku}, {Wanajo},
  {Kiuchi}, {Sekiguchi}, \& {Shibata}}]{Tanaka2014}
{Tanaka}, M., {Hotokezaka}, K., {Kyutoku}, K., {et~al.} 2014, \apj, 780, 31,
  \dodoi{10.1088/0004-637X/780/1/31}

\bibitem[{{Tanaka} {et~al.}(2020){Tanaka}, {Kato}, {Gaigalas}, \&
  {Kawaguchi}}]{Tanaka2020}
{Tanaka}, M., {Kato}, D., {Gaigalas}, G., \& {Kawaguchi}, K. 2020, \mnras, 496,
  1369, \dodoi{10.1093/mnras/staa1576}

\bibitem[{Timmes \& Swesty(2000)}]{Timmes_2000}
Timmes, F.~X., \& Swesty, F.~D. 2000, The Astrophysical Journal Supplement
  Series, 126, 501, \dodoi{10.1086/313304}

\bibitem[{{Wanajo} {et~al.}(2022){Wanajo}, {Fujibayashi}, {Hayashi}, {Kiuchi},
  {Sekiguchi}, \& {Shibata}}]{Wanajo2022}
{Wanajo}, S., {Fujibayashi}, S., {Hayashi}, K., {et~al.} 2022, arXiv e-prints,
  arXiv:2212.04507, \dodoi{10.48550/arXiv.2212.04507}

\bibitem[{{Wanajo} {et~al.}(2014){Wanajo}, {Sekiguchi}, {Nishimura}, {Kiuchi},
  {Kyutoku}, \& {Shibata}}]{Wanajo2014}
{Wanajo}, S., {Sekiguchi}, Y., {Nishimura}, N., {et~al.} 2014, \apjl, 789, L39,
  \dodoi{10.1088/2041-8205/789/2/L39}

\bibitem[{{Wu} {et~al.}(2022){Wu}, {Ricigliano}, {Kashyap}, {Perego}, \&
  {Radice}}]{Wu2022}
{Wu}, Z., {Ricigliano}, G., {Kashyap}, R., {Perego}, A., \& {Radice}, D. 2022,
  \mnras, 512, 328, \dodoi{10.1093/mnras/stac399}

\bibitem[{{Yang} {et~al.}(2018){Yang}, {East}, \& {Lehner}}]{Yang2018}
{Yang}, H., {East}, W.~E., \& {Lehner}, L. 2018, \apj, 856, 110,
  \dodoi{10.3847/1538-4357/aab2b0}

\bibitem[{{Zhu} {et~al.}(2021){Zhu}, {Wu}, {Yang}, {Zhang}, {Yu}, {Gao}, {Cao},
  \& {Liu}}]{Zhu2021}
{Zhu}, J.-P., {Wu}, S., {Yang}, Y.-P., {et~al.} 2021, \apj, 921, 156,
  \dodoi{10.3847/1538-4357/ac19a7}

\end{thebibliography}

\appendix

\section{Light curve calculation}\label{sec:app}

At time $ t = t_f $ -- the end of the simulations, most of the outflow reaches homologous expansion. This enables us to advance the outflow expansion in post-processing using the adiabatic relations. We note that because we do not consider emission from the jet, we can safely ignore the magnetic component which is negligible outside of the jet. The emission calculation is performed on the unbound 3D outflow as follows.
At each line of sight and time, we find the photosphere $ \rph(t,\theta,\phi)=R\left[\tau(t,\rph,\theta,\phi)=1\right] $, and the trapping radius $ \rtr(t,\theta,\phi)=R\left[\tau(t,\rtr,\theta,\phi)=c/v(t_f,\rtr,\theta,\phi)\right] $, where $ v $ is the velocity, and $ \tau $ is the optical depth along a radial line of sight:
\begin{equation}\label{eq:tau}
    \tau(t,r,\theta,\phi) = \left(\frac{t}{t_f}\right)^{-2}\int_{r}^{\infty}\kappa(\theta)\rho(t_f,r',\theta,\phi)(1-v(t_f,r',\theta,\phi)/c)\gamma(t_f,r',\theta,\phi)\mathrm{d}r'\,,
\end{equation}
where $ \rho $ is the mass density.

The photons diffuse out from the trapping radius, such that the luminosity per solid angle in the lab frame due to cooling emission is:
\begin{equation}\label{eq:Lc}
    L_c(t,\theta,\phi) = \left[4p(t_f,\rtr,\theta,\phi)+AQ(t,r,\theta,\phi)\rho(t,r,\theta,\phi)\right]\gamma(t_f,\rtr,\theta,\phi)^2\left(\frac{t_f}{t}\right)^4v(t_f,\rtr,\theta,\phi)\rtr(t,\theta,\phi)^2\,,
\end{equation}
where $ p $ is the thermal pressure, $ \gamma $ is the Lorentz factor, and the heating rate contribution to cooling due to $ \beta $ and $ \gamma $ radioactive decay is \citep{Hotokezaka2016b}:
\begin{equation}\label{eq:Q}
    A = 1-{\rm exp}\left[-\left(\frac{\gamma(t,r,\theta,\phi)}{t_d}\right)^2\right]+0.5;\,\,\,\,\,
    \dot{Q}(t,r,\theta,\phi) = 10^{10}\left[\frac{t_d}{\gamma(t,r,\theta,\phi)}\right]^{-1.3}\,{\rm\frac{erg}{g\,s}}\,,
\end{equation}
and may vary with the composition to within a factor of a few \citep[see e.g.,][]{Freiburghaus1999,Metzger2010a,Korobkin2012,Wanajo2014,Hotokezaka2016b}. The time $ t_d $ is defined as $ t_d \equiv t/{\rm day} $.
The first term in \eqref{eq:Lc} is the adiabatic cooling of the shocked cocoon, as obtained from the radial flux component of the stress-energy tensor. The second term accounts for the radioactive heating of the gas. Both contributions are calculated in the comoving frame, and then boosted to the lab frame.
The luminosity per solid angle due to radioactive decay in the lab frame is:
\begin{equation}\label{eq:Lr}
   L_r(t,\theta,\phi) = \int_{\rtr}^{\infty}A\dot{Q}(t,r,\theta,\phi){\rho(t,r,\theta,\phi)\gamma(t,r,\theta,\phi)^2r^2\mathrm{d}r}\,,
\end{equation}
The total local luminosity is the sum of both contributions, $ L(t,\theta,\phi) = L_c(t,\theta,\phi) +L_r(t,\theta,\phi) $.
The local temperature is calculated at the photosphere and assumed to be blackbody.

To transform the emission from the lab frame luminosity to the observed frame $(\theta_{\rm obs},\phi_{\rm obs})$, we find: 
\begin{equation}\label{eq:tobs}
\tobs(\theta,\phi) = t\left[1-v(t_f,\rtr,\theta,\phi)/c\mu(\theta_{\rm obs},\phi_{\rm obs}\theta,\phi)\right]\,,
\end{equation}
where:
\begin{equation}
    \mu(\theta_{\rm obs},\phi_{\rm obs},\theta,\phi) = {\rm cos}\theta{\rm cos}\theta_{\rm obs}+{\rm sin}\theta_{\rm obs}{\rm sin}\theta{\rm cos}({\phi-\phi_{\rm obs}})\,.
\end{equation}
Then, the observed luminosity is:
\begin{equation}
    L_{\rm obs}(\tobs,\rtr,\theta_{\rm obs},\phi_{\rm obs}) = \int_0^\pi\int_0^{2\pi}D(\tobs,\rtr,\theta,\phi)^4\gamma(\tobs,\rtr,\theta,\phi)L(\tobs,\theta,\phi)\mathrm{d}\mu(\theta_{\rm obs},\phi_{\rm obs},\theta,\phi)\,,
\end{equation}
where $ D(\tobs,\rtr,\theta_{\rm obs},\phi_{\rm obs}) = \left[\gamma(\tobs,\rtr,\theta_{\rm obs},\phi_{\rm obs})\left(1-v(\tobs,\rtr,\theta_{\rm obs},\phi_{\rm obs})\right)\right]^{-1}$ is the Doppler factor of the emitting gas.

The emission peak time is derived as follows. The photons diffuse out from the trapping radius, $ \tau(\rtr) = c/v $, thus Eq.~\eqref{eq:tau} dictates:
\begin{equation}
    \frac{3M\kappa}{4\pi(vt)^2}\gamma(1-v/c) \approx \frac{c}{v}\,.
\end{equation}
Approximating Eq.~\eqref{eq:tobs} to $ \tobs \approx t(1-v/c) $, we find:
\begin{equation}\label{eq:tpeak}
    \tobs \approx \left[\frac{\kappa M \gamma}{4 v c}\left(1-\frac{v}{c}\right)^3\right]^{0.5}~.
\end{equation}

\end{document}